\documentclass[useAMS,usenatbib]{mn2e}
\usepackage{epsfig}

\def\etal{et~al.}
\def\spose#1{\hbox to 0pt{#1\hss}}
\def\lta{\mathrel{\spose{\lower 3pt\hbox{$\mathchar"218$}}
     \raise 2.0pt\hbox{$\mathchar"13C$}}}
\def\gta{\mathrel{\spose{\lower 3pt\hbox{$\mathchar"218$}}
     \raise 2.0pt\hbox{$\mathchar"13E$}}}

\title[]{On the fundamental dichotomy in the local radio-AGN population:
  accretion, evolution, and host galaxy properties}

\author[P.~N.~Best and T.~M.~Heckman]{P.~N.~Best,$^1$\thanks{Email:
    pnb@roe.ac.uk} and T. M. Heckman$^2$
\\
$^1$ SUPA\thanks{Scottish Universities Physics Alliance}, Institute for
Astronomy, Royal Observatory Edinburgh, Blackford Hill, Edinburgh EH9 3HJ \\
$^2$ Department of Physics \& Astronomy, The Johns Hopkins University,
Baltimore, MD 21218, USA\\
}

\begin{document}

\pagerange{\pageref{firstpage}--\pageref{lastpage}}
\pubyear{2011}

\label{firstpage}

\maketitle

\begin{abstract}
\noindent A sample of 18286 radio-loud AGN is presented, constructed by
combining the 7th data release of the Sloan Digital Sky Survey with the
NRAO VLA Sky Survey (NVSS) and the Faint Images of the Radio Sky at Twenty
centimetres (FIRST) survey. Using this sample, the differences between
`high-excitation' (or `quasar-mode'; hereafter HERG) and `low-excitation'
(`radio-mode'; LERG) radio galaxies are investigated. A primary difference
between the two radio source classes is the distinct nature of the
Eddington-scaled accretion rate onto their central black holes: HERGs
typically have accretion rates between one per cent and ten per cent of
their Eddington rate, whereas LERGs predominatly accrete at a rate below
one per cent Eddington. This is consistent with models whereby the
population dichotomy is caused by a switch between radiatively efficient
and radiatively inefficient accretion modes at low accretion rates.  Local
radio luminosity functions are derived separately for the two populations,
for the first time, showing that although LERGs dominate at low radio
luminosity and HERGs begin to take over at $L_{\rm 1.4 GHz} \sim
10^{26}$W\,Hz$^{-1}$, examples of both classes are found at all radio
luminosities. Using the $V/V_{\rm max}$ test it is shown that the two
populations show differential cosmic evolution at fixed radio luminosity:
HERGs evolve strongly at all radio luminosities, while LERGs show weak or
no evolution. This suggests that the luminosity-dependence of the
evolution previously seen in the radio luminosity function is driven, at
least in part, by the changing relative contributions of these two
populations with luminosity. The host galaxies of the radio sources are
also distinct: HERGs are typically of lower stellar mass, with lower black
hole masses, bluer colours, lower concentration indices, and less
pronounced 4000\AA\ breaks indicating younger stellar populations. Even if
samples are matched in radio luminosity and stellar and black hole masses,
significant differences still remain between the accretion rates, stellar
populations, and structural properties of the host galaxies of the two
radio source classes. These results offer strong support to the developing
picture of radio-loud AGN in which HERGs are fuelled at high rates through
radiatively-efficient standard accretion disks by cold gas, perhaps
brought in through mergers and interactions, while LERGs are fuelled via
radiatively inefficient flows at low accretion rates. In this picture, the
gas supplying the LERGs is frequently associated with the hot X-ray haloes
surrounding massive galaxies, groups and clusters, as part of a radio-AGN
feedback loop.
\end{abstract}


\section{Introduction}

Active Galactic Nuclei (AGN) are associated with the accretion of material
onto supermassive black holes, of roughly a million to a billion solar
masses, located near the centres of their host galaxies. Supermassive
black holes are found in essentially all massive galaxies
\citep[e.g.][]{mag98a}, with a mass that correlates strongly with the
stellar mass \citep[e.g.][]{mar03,har04} or velocity dispersion
\citep[e.g.][]{geb00,fer00} of the surrounding galaxy bulge. It is now
widely accepted that the build-up of these supermassive black holes and
that of their host spheroids are intimately linked. Evidence is growing
that AGN activity may play an important role in the evolution of the host
galaxy, with AGN outflows being responsible for controlling or terminating
star-formation \citep[e.g. see review by][]{cat09a}.

AGN activity occurs in at least two different modes, each of which may
have an associated, yet different, feedback effect upon the host
galaxy. The most commonly-considered mode of AGN activity is the
`standard' accretion mode associated with quasars. In this mode, which has
been variously referred to as `quasar-mode', `cold-mode', `radiative
mode', `fast-accretor', `high-excitation', or `strong-lined', material is
accreted onto the black hole through a radiatively--efficient,
optically-thick, geometrically thin accretion disk \citep[e.g.][]{sha73}.
These AGN radiate across a very broad range of the electromagnetic
spectrum \citep[e.g.][]{elv94} although a dusty structure surrounding the
black hole and accretion disk, often referred to as a torus, obscures the
emission at some wavelengths when the AGN is seen edge-on \citep[][and
  references therein]{ant93}. They are often associated with
star-formation activity in the host galaxies \citep[e.g.][]{kau03c},
although possibly with delays between the star formation and the AGN
activity \citep[e.g.][and references therein]{wil10,tad11}.  A fraction of
these AGN are radio-loud, possessing powerful radio jets that can extend
for tens or hundreds of kpc. 

This radiatively-efficient accretion mode may be important in curtailing
star formation at high redshifts and setting up the tight relationship
between black hole and bulge masses observed in the nearby Universe
\citep[e.g.][]{sil98,fab99,kin03,rob06}. Observational evidence that
quasars can accelerate high-velocity winds is plentiful, although there is
much debate as to whether these are thermal `energy-driven' winds, or
`momentum-driven' by radiation pressure \citep[e.g.][and references
  therein]{cat09a}. In radio-loud AGN, the powerful radio jets may also
shock-accelerate the gas \citep[e.g.][]{bes00c}, driving bipolar winds at
speeds up to thousands of km/s \citep[e.g.][]{nes08}. Considerable
uncertainty remains as to the mass and energy content of material that is
driven out by these winds, the size-scale of the outflows in radio-quiet
quasars, and the relative importance of the different mechanisms that may
drive the winds.

There is a second mode of AGN activity, in which the accretion of material
on to the black hole leads to little radiated energy, but can lead to the
production of highly-energetic radio jets. It was first noted by
\citet{hin79} that a population of low luminosity radio sources exist in
which the strong emission lines normally found in powerful AGN were
absent. It has since been shown that these radio sources also exhibit no
accretion-related X-ray emission, nor infrared emission from a putative
torus \citep[e.g.][and references therein]{har07}, and are thus
intrinsically different from the quasar-like AGN. \citet{bes05a} showed
that these low-luminosity radio sources are hosted by fundamentally
different host galaxies to emission-line selected (quasar-like) AGN, in
terms of their stellar masses and host galaxy properties. These AGN, which
have been referred to as `radiatively inefficient', `radio-mode',
`hot-mode', `slow-accretor', `low-excitation', or `weak-lined', are
believed to be fuelled through advection-dominated accretion flows
(ADAFs), which are optically thin, geometrically thick, accretion flows
\citep[e.g.][]{nar95}.  They emit the bulk of their energy in kinetic form
through the radio jets \citep[e.g.][]{mer07}; debate remains as to whether
all of the energy of the jets is associated with accretion, or whether
energy from the spin of the supermassive black hole is also tapped
\citep[][and references therein]{mcn11}.

Although the total cosmic contribution of the energetic output of jets is
nearly two orders of magnitudes lower than that of radiation from the
`quasar-mode' AGN \citep[e.g.][]{cat09b}, the jet energy is all deposited
locally to the system, potentially producing a very efficient feedback
mechanism. This is most directly observed in the bubbles and cavities that
radio-AGN are observed to evacuate in the hot hydrostatic gas haloes of
their host galaxies or surrounding groups and clusters
\citep[e.g.][]{boh93,car94b,mcn00,fab06}. The energies estimated for the
radio sources correlate well with the Bondi accretion rates expected from
the hot gas \citep{all06}; this suggests that this hot gas may form both
the fuel for the radio source, and the repository of its energy, offering
the potential for a feedback cycle.

`Radio-mode' AGN have been widely used in galaxy formation models as a
mechanism to switch off star formation in the most massive galaxies, thus
reproducing both the observed shape of the galaxy luminosity function and
the ``old, red and dead'' nature of massive early-type galaxies
\citep{cro06,bow06}. \citet{bes05b} showed that the prevalence of
radio-AGN activity was a very strong function of the mass of the host
galaxy, rising to over 30\% in the most massive systems. Using a scaling
relation between the radio luminosity and the mechanical energy of the jet
\citep[cf.][]{bir04,cav10}, \citet{bes06a} went on to show that the
time-averaged energetic output of these sources is indeed sufficient to
counter-balance gas cooling in early-type galaxies of all masses.  On a
larger-scale, radio-AGN are almost ubiquitous in the brightest cluster
galaxies of cool-core clusters \citep{bur90,bes07}, and have been invoked
as the solution to both the `cooling flow' and the `entropy floor'
problems in the intra-cluster medium of groups and clusters \citep[][and
  references therein]{mcn07}.

The radiatively inefficient and radiatively efficient AGN clearly have
fundamental differences, but the precise origin of these differences
remains unclear. Some authors have argued that it relates to the origin of
the fuelling gas, with accretion of cold gas leading to a stable accretion
disk and a radiatively efficient accretion, while the accretion of hot gas
via the Bondi mechanism would produce the jet-dominated radiatively
inefficient AGN \citep[e.g.][]{har07}. Others argue that the spin of the
black hole is important \citep{mcn11,mar11}. A third hypothesis is that it
is solely (or primarily) driven by the Eddington-scaled accretion rate
onto the black hole, with the ADAF mode occurring when the accretion rate
is well below the Eddington limit. This was the prediction in the original
work of \citet{nar95}, and support for this picture has come from recent
work indicating that broad-line AGN (ie. quasar-like AGN seen face-on)
have lower limits to their accretion rates at around 1 percent of
Eddington \citep{kol06,tru09a,tru11}, and indications that a switch
between flat-spectrum radio quasars and BL Lac objects (which are believed
to be beamed versions of the radiatively-inefficient sources) also occurs
at that Eddington rate \citep{wu11,ghi11}. This hypothesis is used by
synthesis models for AGN evolution \citep[e.g.][]{mer08} that have been
constructed based upon the two different accretion modes.

A critical input to these AGN evolution models, and to understanding the
evolving feedback role that AGN may play in galaxy evolution
\citep[cf.][]{cro06,bow06} is a full understanding of the different AGN
populations, their distribution in luminosity, their host galaxies, and
their cosmic evolution. The cleanest method for selecting samples of
radiatively inefficient AGN is through radio selection, using the emission
of their jets. Radiatively inefficient AGN are detectable at other
wavelengths: in particular, around 30\% of the population of `X-ray bright
optically normal' galaxies found in X-ray surveys show X-ray spectra with
no absorption, yet no evidence of AGN activity at optical wavelengths
\citep{tru09b}, and are interpreted as being radiatively inefficient AGN
in which the X-rays relate to a beamed component of the jet emission
\citep{har09}. Nevertheless, only at radio wavelengths is the selection
function well-understood, and large samples can be constructed. In
addition, corresponding samples of radiatively efficient (radio-loud) AGN
can be constructed at the same time in exactly the same manner, allowing
direct comparisons between the two.

This paper presents a large sample of radio sources drawn from the Sloan
Digital Sky Survey \citep[SDSS;][]{yor00}, and compares the properties of
radio-selected radiatively efficient and inefficient AGN. The selection of
the sample and the classification of the radio sources is described in
Section~\ref{sec_sample}.  Section~\ref{sec_radprops} investigates the
nature of the radio sources: their accretion rates, their luminosity
function, and their cosmic evolution.  Section~\ref{sec_optprops} compares
various properties of the host galaxies of the two classes of sources. The
results are discussed and conclusions drawn in
Section~\ref{sec_concs}. Throughout the paper, the cosmological parameters
are assumed to have values of $\Omega_m = 0.3$, $\Omega_{\Lambda} = 0.7$,
and $H_0 = 70$\,km\,s$^{-1}$Mpc$^{-1}$.

\section{Sample selection and properties}
\label{sec_sample}

\subsection{The overall radio source sample}
\label{sec_samp1}

The sample of radio sources was constructed by combining the 7th data
release \citep[DR7;][]{aba09} of the SDSS spectroscopic sample with the
National Radio Astronomy Observatory (NRAO) Very Large Array (VLA) Sky
Survey \citep[NVSS;][]{con98} and the Faint Images of the Radio Sky at
Twenty centimetres (FIRST) survey \citep{bec95}, broadly following the
techniques described by \citet{bes05a} for the earlier SDSS DR2 sample.
The parent sample for the DR7 matching is the 927,552 galaxies in the
value-added spectroscopic catalogues produced by the group from the Max
Planck Institute for Astrophysics, and Johns Hopkins University (hereafter
MPA-JHU), and available at http://www.mpa-garching.mpg.de/SDSS/
\citep[cf.][]{bri04b}. These galaxies were cross-matched with the NVSS and
FIRST radio sources following the method of \citet{bes05a}, but adopting
the improvement described by \citet{don09} for identification of sources
without FIRST counterparts. The cross-matching goes down to a flux density
level of 5\,mJy, which means that the sample probes down to radio
luminosities of $L_{\rm 1.4GHz} \approx 10^{23}$W\,Hz$^{-1}$ at redshift
$z=0.1$. The sample of detected radio sources is presented in
Table~\ref{tab_sourcesamp}. 

The next step was separation of the radio-AGN from star-forming
galaxies. This has been improved since the DR2 sample, and now makes use
of an optimal combination of three different methods: the method based on
4000\AA\ break strengths and the ratio of radio luminosity to stellar
mass, used in \citet{bes05a}; a method based on the ratio of radio to
emission line luminosity, similar to that presented in \citet{kau08}; a
standard `BPT' emission-line diagnostic method
\citep{bal81,kau03c}. Appendix~\ref{app_sfagn} provides full details of
the combined method. The resultant classifications are provided in
Table~\ref{tab_sourcesamp}. The radio luminosity functions for
star-forming galaxies and radio-loud AGN separately provide broad
confirmation of the success of the classifications
(cf.\ Section~\ref{sec_radlumfunc}). Further tests have been carried out 
to ensure that none of the results of this paper is dependent upon the 
specific details of the SF-AGN separation method.

For the current paper, analysis is restricted to radio sources within the
`main galaxy sample' \citep{str02}, comprising those galaxies with
magnitudes in the range $14.5 < r < 17.77$, and further restricted to the
redshift range $0.01 < z < 0.3$. Within this sample there are 9,168 radio
sources, of which 7,302 are classified as radio-AGN. The median redshift
of the radio-AGN is $z = 0.16$, and 1,245 are located at redshift $z \le
0.1$, to which redshift range some of the analyses are restricted. 

Properties of the radio source host galaxies are drawn from the value
added catalogues of the MPA-JHU group. In particular, these include total
stellar masses \citep{kau03a}, accurate emission line fluxes, after
subtraction of the modelled stellar continuum to account for underlying
stellar absorption features \citep{tre04}, parameters determined directly
from the spectra such as 4000\AA\ break strengths and galaxy velocity
dispersions \citep[cf. ][]{bri04b}, and a compendium of basic parameters
from the imaging data such as galaxy magnitudes, colours, sizes and
structural parameters \citep[see][for more details]{yor00}. As noted by
the MPA-JHU group, the formal line flux uncertainties quoted in the
MPA-JHU DR7 catalogue significantly underestimate the true values (as
determined by comparing derived line fluxes of sources observed multiple
times), and so the line flux uncertainties have been scaled by the factors
recommended by the MPA-JHU team.

\begin{table*}
\caption{\label{tab_sourcesamp} Properties of the 18286 SDSS radio
  galaxies. Only the first 20 sources are listed here: the full table is
  available electronically. The first three columns give the
  identification of the targetted galaxies through their SDSS plate and
  fibre IDs and the date of the observations. Columns 4 to 6 give the RA,
  Dec and redshift of the galaxies. Column 7 gives the integrated flux
  density of the source as measured using the NVSS. Column 8 provides the
  radio classification of the source, following Best et~al (2005): class 1
  are single--component NVSS sources with a single FIRST match; class 2
  are single--component NVSS sources resolved into multiple components by
  FIRST; class 3 are single--component NVSS sources without a FIRST
  counterpart; class 4 sources are those which have multiple NVSS
  components. Where a galaxy has a central FIRST component, the integrated
  flux density and offset from the optical galaxy of that central FIRST
  component are given in columns 9 and 10. Column 11 provides a flag
  classifying the source as either a radio--loud AGN (1) or a star-forming
  galaxy (0), according to the criteria described in
  Appendix~\ref{app_sfagn}. Column 12 indicates whether the source is
  included in the full statistical ``main sample'' studied in this paper
  (ie. SDSS main sample target with $14/5 \le r \le 17.77$). Columns 13
  and 14 indicate whether selected radio-loud AGN are classified as LERGs
  or HERGs, respectively (if such classification is possible; sources with
  0 for both cases are unclassifiable using current data).}
\begin{center}
\begin{tabular}{cccccccccccccc}
\hline
Plate & Julian & Fibre &      RA      &      Dec     & z &
$S_{\rm NVSS}$ & Radio & $S_{\rm FIRST}$ & Offset & AGN & Main &LERG&
HERG\\

 ID   &  Date  &  ID   & \multicolumn{2}{c}{(J2000)} &   &
1.4\,GHz       & Class &  1.4\,GHz       &        &  &  Samp   \\

      &        &       &    (hr)      &    (deg)     &   & 
(Jy)           &       &    (Jy)         & ($''$) & & \\
\hline
 266 & 51602 &   5 &  9.784226 &  -0.81043 &  0.4486 & 0.0069 &  1 & 0.0042 &   1.76 &  1 &  0 &  0 &  0 \\
 266 & 51602 &  26 &  9.797071 &  -0.34230 &  0.1348 & 0.0963 &  1 & 0.1010 &   1.26 &  1 &  1 &  1 &  0 \\
 266 & 51602 & 100 &  9.742905 &  -0.74164 &  0.2038 & 0.0068 &  1 & 0.0025 &   0.46 &  1 &  1 &  1 &  0 \\
 266 & 51602 & 109 &  9.782474 &  -0.25218 &  0.1304 & 0.0075 &  1 & 0.0043 &   0.51 &  1 &  1 &  0 &  0 \\
 266 & 51602 & 134 &  9.720425 &  -0.54706 &  0.3679 & 0.0091 &  2 & 0.0000 &        &  1 &  0 &  0 &  0 \\
 266 & 51602 & 150 &  9.758252 &  -0.36839 &  0.0530 & 0.0104 &  1 & 0.0010 &   2.02 &  0 &  1 &  0 &  0 \\
 266 & 51602 & 179 &  9.746165 &  -0.50828 &  0.3693 & 0.0059 &  1 & 0.0053 &   0.16 &  1 &  0 &  0 &  0 \\
 266 & 51602 & 235 &  9.706746 &  -0.00139 &  0.1459 & 0.0054 &  1 & 0.0049 &   0.43 &  0 &  1 &  0 &  0 \\
 266 & 51602 & 439 &  9.720012 &   0.41417 &  0.0252 & 0.0081 &  3 & 0.0000 &        &  0 &  0 &  0 &  0 \\
 266 & 51602 & 504 &  9.764199 &   0.63871 &  0.0303 & 0.0052 &  1 & 0.0028 &   2.09 &  0 &  1 &  0 &  0 \\
 266 & 51602 & 507 &  9.758241 &   0.25554 &  0.1291 & 0.0275 &  1 & 0.0269 &   0.32 &  1 &  0 &  1 &  0 \\
 266 & 51602 & 550 &  9.776251 &   0.46721 &  0.4505 & 0.0088 &  1 & 0.0058 &   0.44 &  1 &  0 &  0 &  0 \\
 266 & 51602 & 554 &  9.787119 &   0.66564 &  0.0201 & 0.0180 &  1 & 0.0131 &   0.79 &  1 &  1 &  1 &  0 \\
 266 & 51602 & 559 &  9.786605 &   0.70274 &  0.0305 & 0.0063 &  1 & 0.0045 &   2.72 &  0 &  1 &  0 &  0 \\
 266 & 51602 & 577 &  9.785432 &   0.73798 &  0.2616 & 0.0489 &  1 & 0.0094 &   0.78 &  1 &  1 &  1 &  0 \\
 266 & 51602 & 617 &  9.805367 &   0.78802 &  0.2112 & 0.0082 &  1 & 0.0078 &   0.66 &  1 &  1 &  0 &  0 \\
 266 & 51630 & 361 &  9.706815 &   1.14969 &  0.4498 & 0.0523 &  1 & 0.0454 &   0.56 &  1 &  0 &  1 &  0 \\
 266 & 51630 & 529 &  9.772283 &   1.08112 &  0.5768 & 0.0146 &  1 & 0.0112 &   0.41 &  1 &  0 &  0 &  0 \\
 267 & 51608 &  19 &  9.907105 &  -0.92869 &  0.3583 & 0.1848 &  1 & 0.1839 &   0.77 &  1 &  0 &  1 &  0 \\
 267 & 51608 &  34 &  9.944658 &  -0.02334 &  0.1391 & 0.1660 &  4 & 0.0022 &   0.27 &  1 &  1 &  1 &  0 \\
 ... &  ...  & ... &     ...   &     ...   &    ...  &   ...  &... &   ...  &   ...  & ...&... & ...& ...\\ 
 ... &  ...  & ... &     ...   &     ...   &    ...  &   ...  &... &   ...  &   ...  & ...&... & ...& ...\\ 
\hline
\end{tabular}
\end{center}
\end{table*}

\subsection{High/low excitation classification of the sample}
\label{sec_herglerg}

A key requirement of the current analysis is an ability to separate the
radio source sample into the two fundamentally-different AGN classes. For
consistency with previous works on radio source samples, the nomenclature
of ``high-excitation'' and ``low-excitation'' radio sources (HERGs and
LERGs) will be adopted in this paper to denote these.

The work of \citet{lai94}, based on very powerful radio-AGN from the 3CR
sample, suggested a fairly pronounced division between the two
classes. Laing et~al.\ classified as HERGs those sources which had the
line flux ratio [OIII]~5007 / H$\alpha > 0.2$ and an equivalent width of
the [OIII] line EW$_{\rm [OIII]} > 3$\AA\ (where emission line equivalent
widths are assigned positive values). \citet{tad98} similarly found that
low-excitation sources in the 2Jy radio sample, with EW$_{\rm [OIII]} <
10$\AA, stand out from other galaxies in both their [OIII]~5007 /
[OII]~3727 line ratio and the ratio of emission line to radio luminosity.
Lower luminosity radio samples, however, show a less-clear division
between the two classes, since the emission line luminosities of the
high-excitation sources correlate strongly with radio luminosity
\citep{raw91b} and so become much weaker (and less easy to distinguish
from LERG lines) in low-luminosity systems \citep[cf.][]{zir95,kau08}.

More recent analyses separating HERGs and LERGS have been carried out for
large samples of radio galaxies \citep[e.g.][]{but10,cid10,bal10}, largely
based on emission line diagnostics to separate Seyfert from LINER galaxies
devised by \citet{kew06}. Note that a growing literature of work
\citep[e.g.][and references therein]{cid11} indicates that a significant
proportion of LINERs are not due to AGN activity, but rather the emission
lines are photo-ionised by post-asymptotic giant branch (post-AGB) stars
in old galaxies: such cases can be identified as having an equivalent
width of the H$\alpha$ line below 3\AA\ \citep{cid11}. In the radio
galaxies studied here, the radio activity confirms that an AGN must be
present, but this does not rule out the possibility that the weak emission
lines could still have a post-AGB origin.  

\citet{but10} have defined an ``excitation index'' parameter, combining
four emission line ratios: $EI = \rm{log}_{10}(\rm{[OIII]}/H\beta) -
\frac{1}{3}\left[\rm{log}_{10}({\rm [NII]} / H\alpha) +
  \rm{log}_{10}(\rm{[SII]} / H\alpha) + \rm{log}_{10}(\rm{[OI]} /
  H\alpha)\right]$. They demonstrate this parameter to be bimodal and use
it to classify the galaxies, dividing the LERG and HERG populations at a
value of $EI=0.95$. For the SDSS radio galaxy sample defined in
Section~\ref{sec_samp1}, in many cases the full set of emission lines
needed to classify the host galaxies via the Excitation Index are either
not detected, or the signal-to-noise of the detections is low. Therefore a
multiple approach was adopted to carry out the classifications, working
down the following series of possibilities for each source until a
classification was derived.

\begin{enumerate}
\item If all six emission lines were detected and the excitation index was
  at least 1$\sigma$ away from 0.95, the radio source was classified using
  the excitation index [783 LERG; 95 HERG].
\item If four lines were reliably detected for one of the individual
  \citet{kew06} diagnostic diagrams and the source lay at least 1$\sigma$
  from the division line, then that diagnostic diagram was used for
  classification [330 LERG; 2 HERG].
\item If the equivalent width of the [OIII] emission line was at least
  1$\sigma$ above 5\AA\ then the source was classified as high excitation
  [65 HERG].
\item Classification options (i)--(iii) were repeated but with the
  1$\sigma$ criterion removed [305 LERG; 30 HERG].
\item If [NII] and H$\alpha$ emission line measurements were available for
  the source, then the [NII]/H$\alpha$ vs [OIII]/H$\alpha$ emission line
  diagnostic of \citet{cid10} was used, using either a detection or a 
  limit (if definitive) for the [OIII] emission line [769 LERG; 24 HERG]
\end{enumerate}

\begin{figure}
\centerline{
\psfig{file=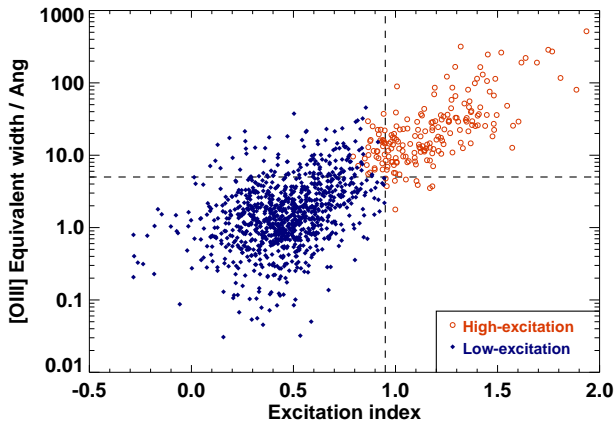,width=8.6cm,clip=} 
}
\caption{\label{fig_exind} The distribution of the HERG/LERG-classified
radio sources on the [OIII] equivalent width versus excitation index
\citep{but10} plane, for galaxies with both parameters measured. This
demonstrates the broad consistency of the two main approaches used to
classify the radio sources.}
\end{figure}

To illustrate the consistency of these cuts, Figure~\ref{fig_exind} shows
the distribution of the classified galaxies on the EW$_{\rm [OIII]}$ {\it
  vs} $EI$ plane, for galaxies with both parameters measured.  It is clear
that the radio sources do show a good correlation and that both methods
can provide suitable classifications. There is a small population of LERGs
with relatively high [OIII] equivalent widths but low excitation indices,
and therefore a risk that classification method (iii) may lead to
contamination of the HERG sample by such sources. However, such
contamination is expected to be small: of the 65 sources classified by
method (iii), 53 have excitation index measurements within 1$\sigma$ of
0.95, and all of the remainder have a limiting [OIII]/H$\alpha \gta 1$,
strongly suggesting they are indeed HERGs.

\begin{figure*}
\centerline{
\begin{tabular}{cc}
\psfig{file=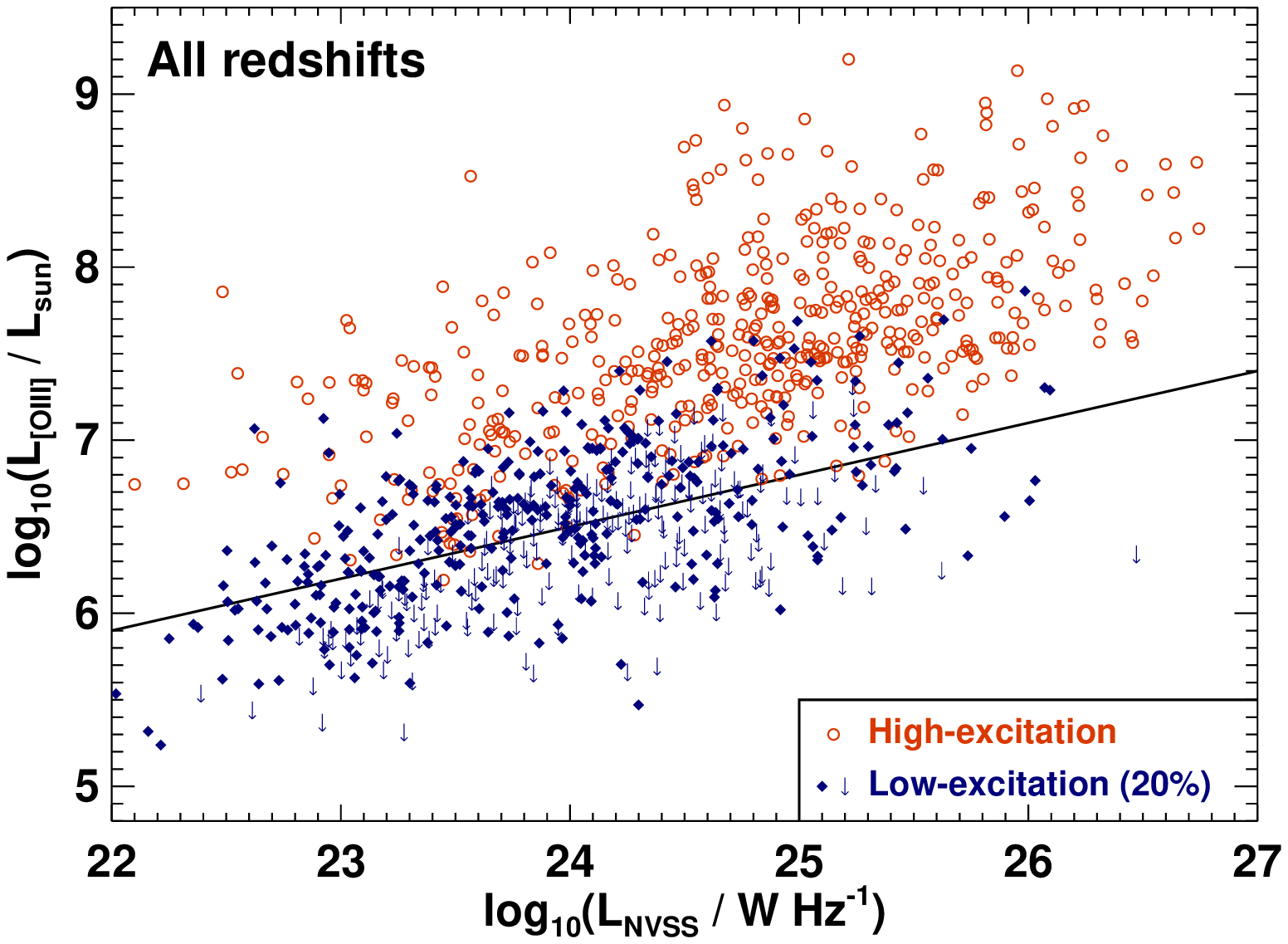,width=8.6cm,clip=} &
\psfig{file=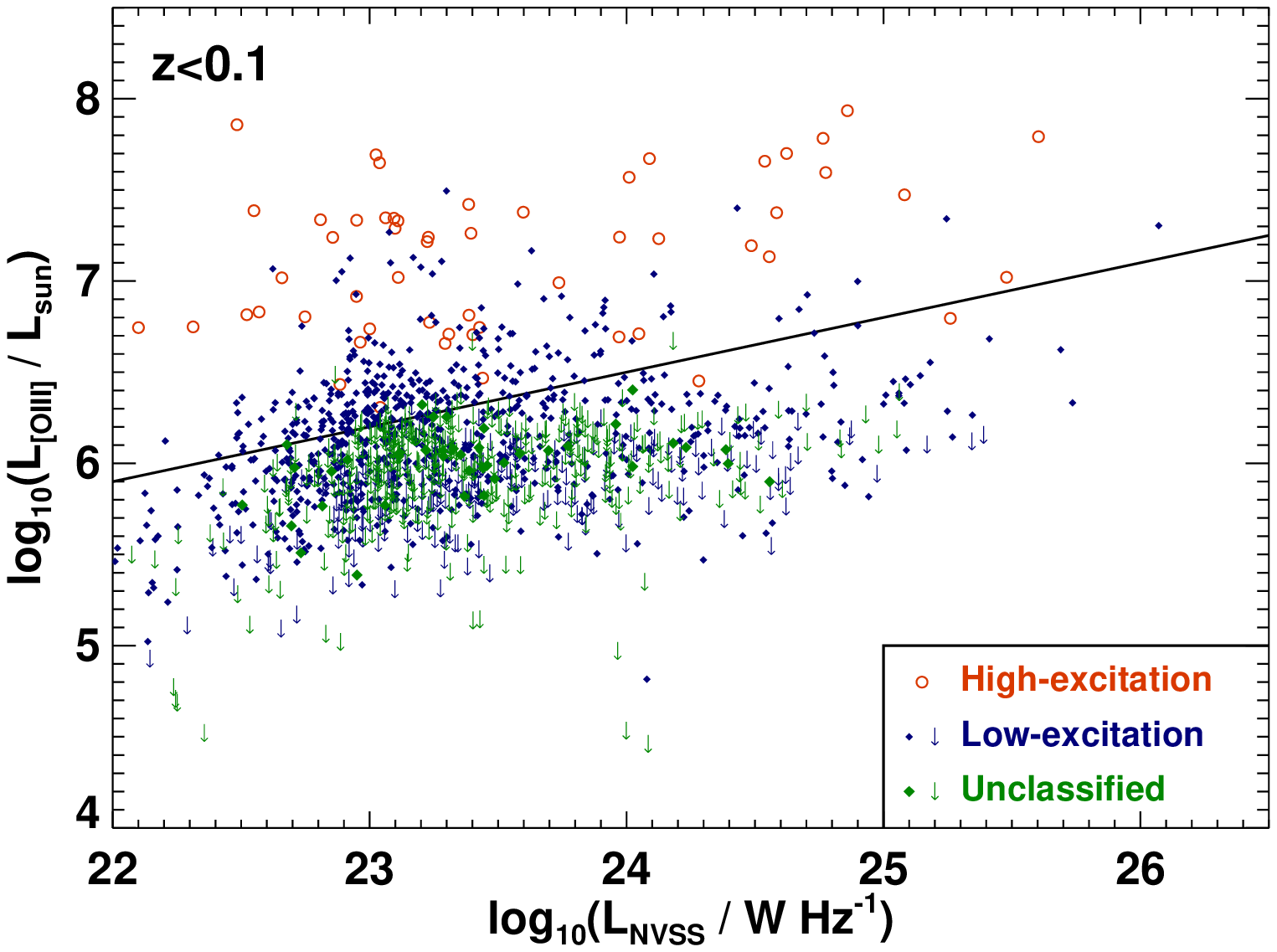,width=8.6cm,clip=}
\end{tabular}
}
\caption{\label{fig_loptlrad}{\it Left:} the [OIII] emission line
  luminosity versus radio luminosity for HERG/LERG-classified radio
  sources. The solid line indicates an approximate lower limit to the
  distribution of the HERGs, below which unclassified sources can be
  classified as LERGs with reasonable confidence. Note that only 20\%
  (randomly selected) of LERGs are plotted to avoid over-crowding of the
  figure. {\it Right:} the same plot, but only for radio sources with $z <
  0.1$, and also including the unclassified galaxies. It can be seen that
  all but four of the unclassified galaxies in this redshift range can be
  robustly classified as LERGs using the criterion derived from the left
  panel.}
\end{figure*}

Despite using all of these different mechanisms, only about a third of the
radio sources were able to be classified. Many are simply undetected in
emission lines, suggesting that they are likely to be LERGs -- although in
some cases this may just be due to the relative faintness of the
source. To investigate a possible mechanism for classifying these sources,
the left panel of Figure~\ref{fig_loptlrad} shows the distribution of
[OIII] line luminosity versus radio luminosity for the classified
sources. As discussed above, the two populations do occupy different
regions of this plane\footnote{Radio-quiet quasars would lie above and to
  the left of the HERG sources \citep[e.g.][]{xu99}.}, although there is
significant overlap. The solid line in the figure represents an
approximate lower limit of the distribution of the HERGs, and so the 3883
sources with emission line luminosities, or limits, below this can
therefore be fairly securely classified as LERGs.  As demonstrated in the
right panel of Figure~\ref{fig_loptlrad}, this allows robust
classification of all but four of the $z<0.1$ subsample of radio galaxies.

\section{Fundamental properties of the radio sources}
\label{sec_radprops}

\subsection{Local radio luminosity functions of HERGs and LERGS}
\label{sec_radlumfunc}

\begin{table*}
\caption{\label{tab_lumfunc} The local radio luminosity functions at
  1.4\,GHz, derived separately for the HERG and LERG populations. The
  first column shows the range of 1.4\,GHz radio luminosities considered
  in each bin. The second and third columns show the total number of radio
  sources and the space density of these, in units of number per log$_{10}
  L$ per Mpc$^3$, detected out to $z=0.3$. Columns 4 to 7 show the radio
  sources split into star-forming galaxies and radio-loud AGN.
  The eighth column gives the maximum redshift considered for the
  LERG/HERG analysis, in order to minimise the number of unclassified
  sources. The numbers and space densities of LERGs, HERGs, and
  unclassified sources, respectively, are given in columns 9 to
  14. Uncertainties are statistical Poissonian uncertainties only. Note
  that the unclassified sources have negligible contribution compared to
  the LERG population, but for some bins of luminosity they would make a
  significant additional contribution if added to the HERG sample (see
  also Figure~\ref{fig_lumfunc}).}
\begin{tabular}{crcrcrccrcrcrc}
\hline
${\rm log} L_{\rm 1.4 GHz}$&
\multicolumn{2}{c}{All radio sources} &
\multicolumn{2}{c}{Star-forming} &
\multicolumn{2}{c}{Radio-AGN} &&
\multicolumn{2}{c}{LERGs} &
\multicolumn{2}{c}{HERGs} &
\multicolumn{2}{c}{Unclassified} 
\\
W Hz$^{-1}$  &
N~ & ${\rm log}_{10} \rho$ &
N~ & ${\rm log}_{10} \rho$ &
N~ & ${\rm log}_{10} \rho$ &
z$_{\rm max}$&
N~ & ${\rm log}_{10} \rho$ &
N & ${\rm log}_{10} \rho$ &
N & ${\rm log}_{10} \rho$ \\
\hline
22.0-22.3 &  297 & -$3.09^{+0.03}_{-0.03}$ & 284 & -$3.11^{+0.03}_{-0.04}$  &   13 & -$4.44^{+0.15}_{-0.23}$ & 0.10 &  12 & -$4.46^{+0.16}_{-0.25}$ &  1 & -$5.74^{+0.25}$ &    & \\
22.3-22.6 &  385 & -$3.49^{+0.02}_{-0.03}$ & 357 & -$3.54^{+0.03}_{-0.03}$  &   28 & -$4.45^{+0.05}_{-0.05}$ & 0.10 &  24 & -$4.51^{+0.05}_{-0.06}$ &  4 & -$5.64^{+0.19}_{-0.33}$ &    & \\
22.6-22.9 &  532 & -$3.87^{+0.02}_{-0.02}$ & 388 & -$4.00^{+0.02}_{-0.03}$  &  144 & -$4.45^{+0.04}_{-0.04}$ & 0.10 & 138 & -$4.46^{+0.04}_{-0.05}$ &  5 & -$6.01^{+0.16}_{-0.26}$ &  1 & -$6.72$ \\
22.9-23.2 &  674 & -$4.22^{+0.02}_{-0.02}$ & 298 & -$4.57^{+0.03}_{-0.03}$  &  376 & -$4.48^{+0.02}_{-0.03}$ & 0.10 & 339 & -$4.49^{+0.02}_{-0.03}$ &  9 & -$6.09^{+0.13}_{-0.18}$ &    &   \\
23.2-23.5 &  882 & -$4.56^{+0.02}_{-0.02}$ & 221 & -$5.13^{+0.03}_{-0.03}$  &  661 & -$4.69^{+0.02}_{-0.02}$ & 0.10 & 248 & -$4.73^{+0.03}_{-0.03}$ & 12 & -$6.09^{+0.11}_{-0.15}$ &  3 & -$6.66$ \\
23.5-23.8 & 1358 & -$4.75^{+0.01}_{-0.01}$ & 126 & -$5.69^{+0.04}_{-0.05}$  & 1232 & -$4.80^{+0.01}_{-0.01}$ & 0.13 & 377 & -$4.91^{+0.02}_{-0.02}$ & 12 & -$6.42^{+0.11}_{-0.15}$ & 10 & -$6.48$\\
23.8-24.1 & 1615 & -$4.90^{+0.02}_{-0.02}$ &  52 & -$6.35^{+0.06}_{-0.08}$  & 1563 & -$4.91^{+0.02}_{-0.02}$ & 0.15 & 454 & -$4.91^{+0.07}_{-0.09}$ & 10 & -$6.69^{+0.12}_{-0.17}$ &  8 & -$6.77$\\
24.1-24.4 & 1327 & -$5.08^{+0.01}_{-0.01}$ &  19 & -$6.83^{+0.11}_{-0.15}$  & 1308 & -$5.09^{+0.01}_{-0.02}$ & 0.17 & 427 & -$5.19^{+0.02}_{-0.02}$ & 15 & -$6.62^{+0.10}_{-0.14}$ & 14 & -$6.70$\\
24.4-24.7 &  949 & -$5.25^{+0.02}_{-0.02}$ &   3 & -$7.43^{+0.23}_{-0.51}$  &  946 & -$5.26^{+0.02}_{-0.02}$ & 0.17 & 275 & -$5.38^{+0.03}_{-0.03}$ & 15 & -$6.61^{+0.10}_{-0.13}$ &  1 & -$7.85$\\
24.7-25.0 &  561 & -$5.54^{+0.02}_{-0.02}$ &   0 &  ---                   &  561 & -$5.54^{+0.02}_{-0.02}$ & 0.20 & 228 & -$5.66^{+0.03}_{-0.03}$ & 16 & -$6.70^{+0.10}_{-0.13}$ &    &  \\
25.0-25.3 &  303 & -$5.82^{+0.03}_{-0.03}$ &   0 &  ---                   &  303 & -$5.82^{+0.03}_{-0.03}$ & 0.25 & 206 & -$5.91^{+0.03}_{-0.04}$ & 21 & -$6.76^{+0.09}_{-0.12}$ &  5 & -$7.63$\\
25.3-25.6 &  103 & -$6.32^{+0.05}_{-0.06}$ &   0 &  ---                   &  103 & -$6.32^{+0.05}_{-0.06}$ & 0.25 &  57 & -$6.44^{+0.06}_{-0.07}$ & 13 & -$6.98^{+0.09}_{-0.11}$ &    &    \\
25.6-25.9 &   47 & -$6.58^{+0.07}_{-0.08}$ &   0 &  ---                   &   47 & -$6.58^{+0.07}_{-0.08}$ & 0.30 &  29 & -$6.82^{+0.08}_{-0.10}$ & 17 & -$6.95^{+0.10}_{-0.14}$ &  1 & -$8.55$\\
25.9-26.2 &   12 & -$7.18^{+0.12}_{-0.17}$ &   0 &  ---                   &   12 & -$7.18^{+0.12}_{-0.17}$ & 0.30 &   9 & -$7.41^{+0.13}_{-0.19}$ &  3 & -$7.64^{+0.17}_{-0.28}$ &    & \\
26.2-26.5 &    3 & -$7.78^{+0.21}_{-0.43}$ &   0 &  ---                   &    3 & -$7.78^{+0.21}_{-0.43}$ & 0.30 &   0 &   ---                  &  3 & -$7.78^{+0.21}_{-0.43}$ &    & \\
\hline  
\end{tabular}
\end{table*}

\begin{figure}
\centerline{
\psfig{file=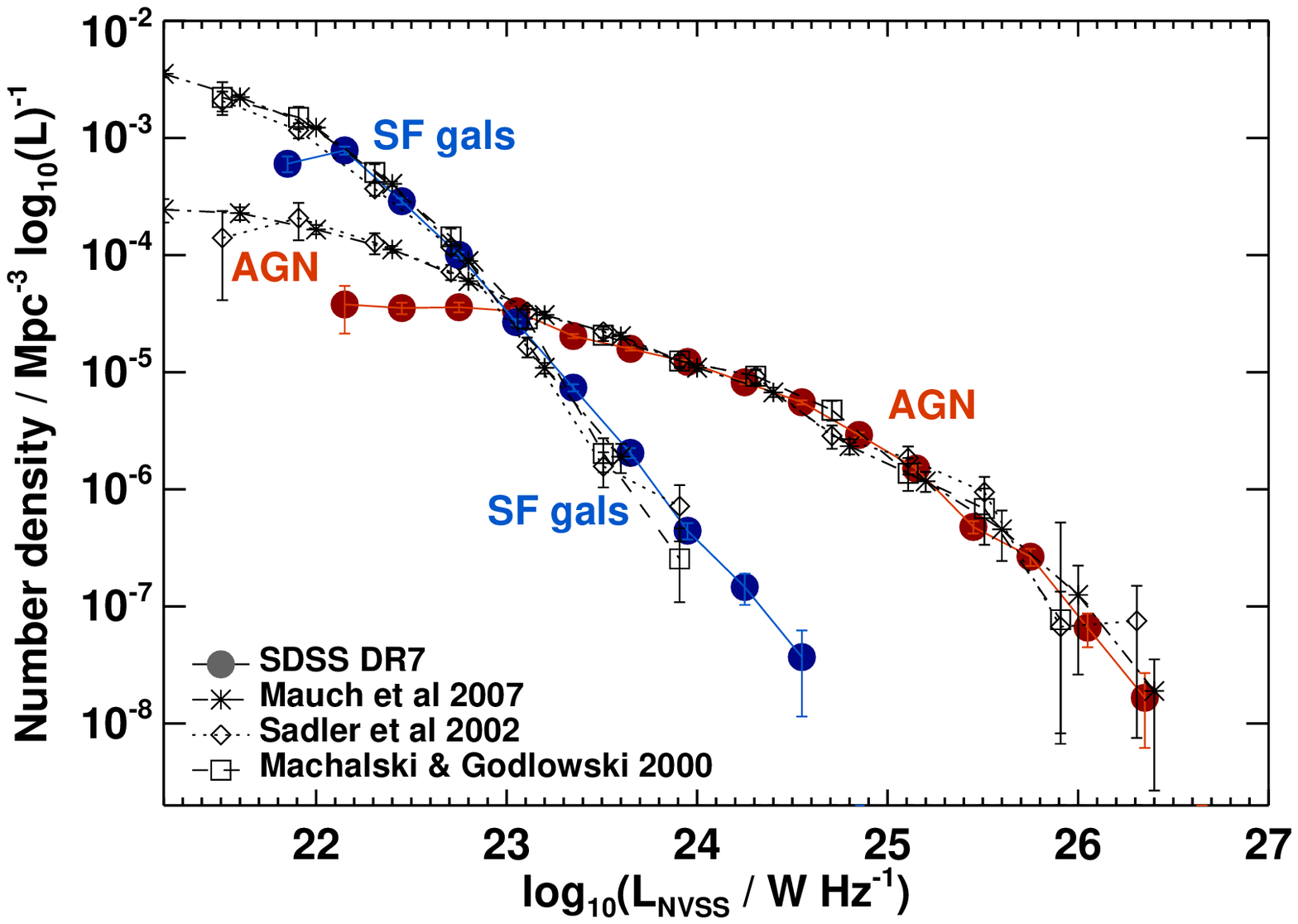,width=8.6cm,clip=} 
}
\caption{\label{fig_rlfsfagn} The local radio luminosity function at
  1.4\,GHz derived separately for radio--loud AGN and star--forming
  galaxies. Filled points connected by solid lines indicate the data
  derived in this paper. For comparison, the results of Machalski \& Godlowski
  (2000) using the Las Campanas Redshift Survey, Sadler
  \etal\ (2002) using the 2-degree field Galaxy Redshift Survey (2dFGRS), 
  and Mauch \etal\ (2007) using the 6-degree field Galaxy Survey (6dFGS) 
  are shown.}
\end{figure}

Radio luminosity functions were calculated in the standard way, as $\rho =
\sum_i 1/V_i$ \citep{sch68b,con89}, where $V_i$ is the volume within which
source $i$ could be detected. This is calculated as $V_i = V_{\rm max} -
V_{\rm min}$, where $V_{\rm max}$ and $V_{\rm min}$ are the volumes
enclosed within the observed sky area out to the upper and lower redshift
limits, respectively, at which each source would be included in the
sample. Redshift limits were determined by the joint radio and optical
selection criteria, namely a radio cut--off of 5\,mJy and optical
cut--offs of $14.5 < r < 17.77$, as well as any imposed redshift limit for
the analysis (e.g. $0.01 < z < 0.3$). The sky area of the overlapping
region between the SDSS DR7 spectroscopic survey and the FIRST radio
survey, after removal of noisy regions around very powerful radio sources,
was calculated to be 2.17 steradians. The summed radio luminosity function
of all radio sources, together with its separation into star-forming
galaxies and radio-loud AGN, are provided in Table~\ref{tab_lumfunc} and
shown in Figure~\ref{fig_rlfsfagn}, and are in excellent agreement with
previous determinations \citep{mac00,sad02,bes05a,mau07}.  Uncertainties
quoted are the statistical Poissonian errors only; at some luminosities
these are so small that they will be under-estimates, with systematic
errors dominating.

\begin{figure}
\centerline{
\psfig{file=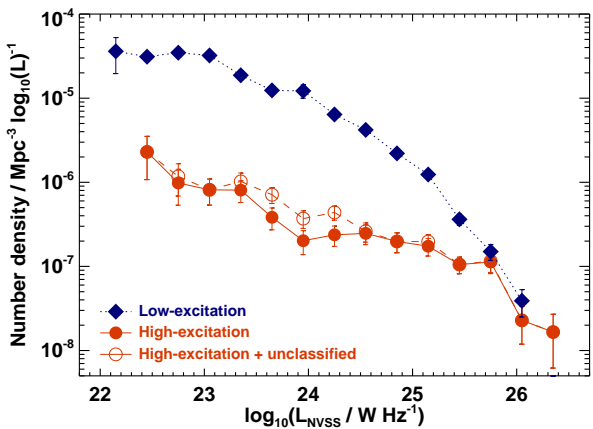,width=8.6cm,clip=} 
}
\caption{\label{fig_lumfunc}The local radio luminosity function at
1.4\,GHz, derived separately for the HERG and LERG populations. Note how
both populations are found across the full range of radio luminosities
studied.}
\end{figure}

Local radio luminosity functions were derived individually for LERGs and
HERGs; these are also tabulated in Table~\ref{tab_lumfunc} and are shown
in Figure~\ref{fig_lumfunc}. In order to minimise the potential influence
of unclassified sources (typically found at higher redshifts) and yet
retain sufficient volume for the rarer luminous sources, the upper limit
of the redshift range used to calculate the radio source space density was
increased with increasing radio luminosity, as indicated in
Table~\ref{tab_lumfunc}. Figure~\ref{fig_lumfunc} illustrates the maximum
effect that the residual unclassified sources could have on the HERG
luminosity function, even if all of them were HERGs: although there are
small changes to some data points, the overall interpretation and
conclusions are unaffected. The potential influence of unclassified
sources on the radio luminosity function of the LERGs is completely
negligible.

Figure~\ref{fig_lumfunc} shows that, as expected, LERGs dominate the radio
source population at relatively low radio luminosities, while the HERGs
begin to dominate at the highest luminosities, beyond $P_{\rm 1.4GHz} \sim
10^{26}$W\,Hz$^{-1}$. However, what is clear (and goes against standard
simplifying assumptions in the literature) is that both populations are
found across the full range of radio luminosities studied: even at radio
luminosities around $P_{\rm 1.4GHz} \sim 10^{23}$W\,Hz$^{-1}$ the HERGs
constitute a few percent of the overall radio-loud AGN population\footnote
{Note that although there may be some contamination of the low-luminosity
  HERG population by radio-quiet quasars or Seyfert galaxies, the bulk of
  these are expected to be genuine radio-loud AGN: as detailed in
  Appendix~A, the separation of star-forming galaxies from AGN was
  designed to exclude the radio-quiet quasars and Seyfert galaxies from
  the AGN category.}.

\subsection{The cosmic evolution of HERGs and LERGS}
\label{sec_cosevol}

\begin{figure}
\centerline{
\psfig{file=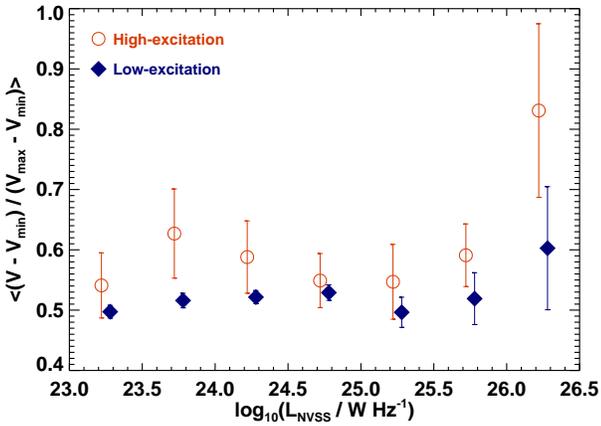,width=8.6cm,clip=}
}
\caption{\label{fig_vvmax}The cosmic evolution, as a function of radio
  luminosity, of the HERG and LERG populations separately, as demonstrated
  using the V/V$_{\rm max}$ test. }
\end{figure}

The cosmic evolution of the HERG and LERG populations can be individually
investigated by using the $V/V_{\rm max}$ test. Taking account of the
lower and upper redshift limits (see Section~\ref{sec_radlumfunc}), the
value of $(V-V_{\rm min}) / (V_{\rm max} - V_{\rm min})$ was calculated
for each source. These were then averaged for HERGs and LERGs separately
within radio luminosity bins; once again, to minimise the effect of
unclassified sources, a varying upper redshift limit was imposed for the
analysis in each radio luminosity bin. The mean values are presented in
Table~\ref{tab_cosevol} and displayed in Figure~\ref{fig_vvmax}.

The HERG population displays clear evidence for cosmic evolution at all
radio luminosities studied, in the sense of there being a tendency for the
sources to be located at larger distances than the median\footnote{Many
  HERGs are classical-double radio sources, sometimes with no detected
  radio core. The possibility exists that sources without cores might be
  missed by the cross-matching procedure at low redshifts due to the large
  angular separation of the components. This would mean that $z_{\rm min}$
  should be higher than estimated, and could lead to an upward bias in
  $(V-V_{\rm min}) / (V_{\rm max} - V_{\rm min})$. In order to ensure that
  this is not the origin of the observed evolution of the HERGs, all HERGs
  with no detected radio core were examined to determine the lowest
  redshift at which the cross-matching procedure would have included them
  in the sample, under the worst-case scenario that no additional FIRST
  components would be detected. For only three sources would this lead to
  a change in the estimate of $z_{\rm min}$ and these each lead to a
  change in the relevant $(V-V_{\rm min}) / (V_{\rm max} - V_{\rm min})$
  value of significantly less than 0.01, which is negligible compared to
  the associated errors.}. In contrast, the LERG population is broadly
consistent with a mean $V/V_{\rm max} \simeq 0.5$ (with the possible
exception of the highest radio luminosity bin). The LERGs therefore show
little or no evidence for any cosmic evolution. Combined with the results
of Section~\ref{sec_radlumfunc}, this result has important implications
for understanding the evolution of the radio luminosity function as a
whole, whereby the differential cosmic evolution seen between powerful and
less powerful radio sources \citep{lon66} may be driven by the switch in
the dominant population with radio luminosity
(cf. Figure~\ref{fig_lumfunc}). At high radio luminosities the HERGs
dominate and, being a strongly-evolving population, lead to strong
evolution of the overall radio source space density at these luminosities
\citep[a factor $\sim$thousand increase in space density out to redshift
  2--3; cf.][and references therein]{dun90,rig11}. In contrast, at low
radio luminosities the LERGs dominate the population, leading to the weak
cosmic evolution seen in the low-luminosity radio population \citep[a
  factor 1.5--2 increase in space density out to $z \sim 0.5$,
  e.g.][]{sad07,don09}. Indeed, even this evolution may be driven to a
large extent by the evolution of the HERGs: locally these contribute $\lta
10$\% of the population at low luminosities, but if they increase in space
density by an order of magnitude out to $z \approx 0.5$ (as they do at
higher radio luminosities) then they would become comparable in numbers to
the LERGs and lead to the observed doubling of the overall space density.

\begin{table}
\begin{center}
\caption{\label{tab_cosevol} V/V$_{\rm max}$ determinations for the LERG
  and HERG populations separately, in various ranges of radio luminosity.
  In each luminosity range, an upper redshift limit (z$_{\rm max}$) is
  defined for the analysis (and given in column 2) in order to reduce the
  number of unclassified sources to negligible levels.}
\begin{tabular}{cccc}
\hline
${\rm log}_{10} L_{\rm 1.4 GHz}$&
z$_{\rm max}$ & 
\multicolumn{2}{c}{$\left<(V - V_{\rm min}) / (V_{\rm max} - V_{\rm min}) \right>$} \\
W Hz$^{-1}$ & & LERG & HERG \\
\hline
23.0-23.5 & 0.13 & 0.50$\pm$0.01 & 0.54$\pm$0.05  \\
23.5-24.0 & 0.13 & 0.51$\pm$0.01 & 0.63$\pm$0.07  \\
24.0-24.5 & 0.17 & 0.52$\pm$0.01 & 0.59$\pm$0.06  \\
24.5-25.0 & 0.20 & 0.53$\pm$0.01 & 0.55$\pm$0.04  \\
25.0-25.5 & 0.20 & 0.50$\pm$0.02 & 0.55$\pm$0.06  \\
25.5-26.0 & 0.30 & 0.52$\pm$0.04 & 0.59$\pm$0.05  \\
26.0-26.5 & 0.30 & 0.60$\pm$0.10 & 0.83$\pm$0.14  \\
\hline
\end{tabular}
\end{center}
\end{table}

\subsection{Eddington-scaled accretion rates of HERGs and LERGS}
\label{sec_eddrate}

As discussed in the introduction, a popular hypothesis for the difference
between LERGs and HERGs relates to the Eddington-scaled accretion rates on
to the black hole. The Eddington-scaled accretion rate can be estimated
for the radio sources by comparing the total energetic output of the black
hole, calculated as the sum of the radiative luminosity and the jet
mechanical luminosity, with the Eddington luminosity.

The bolometric radiative luminosity of each radio source was estimated
from the observed luminosity of the [OIII]~5007 emission line, using the
relation determined by \citet{hec04}: $L_{\rm rad} = 3500 L_{\rm
  OIII}$. Where the [OIII] line was not detected, an upper limit to the
radiative luminosity was set instead. The uncertainty on individual
estimates of the bolometric radiative luminosity is $\approx 0.4$ dex,
from the scatter around the $L_{\rm rad}$ {\it vs} $L_{\rm OIII}$ relation
\citep{hec04}.

\begin{figure}
\centerline{
  \psfig{file=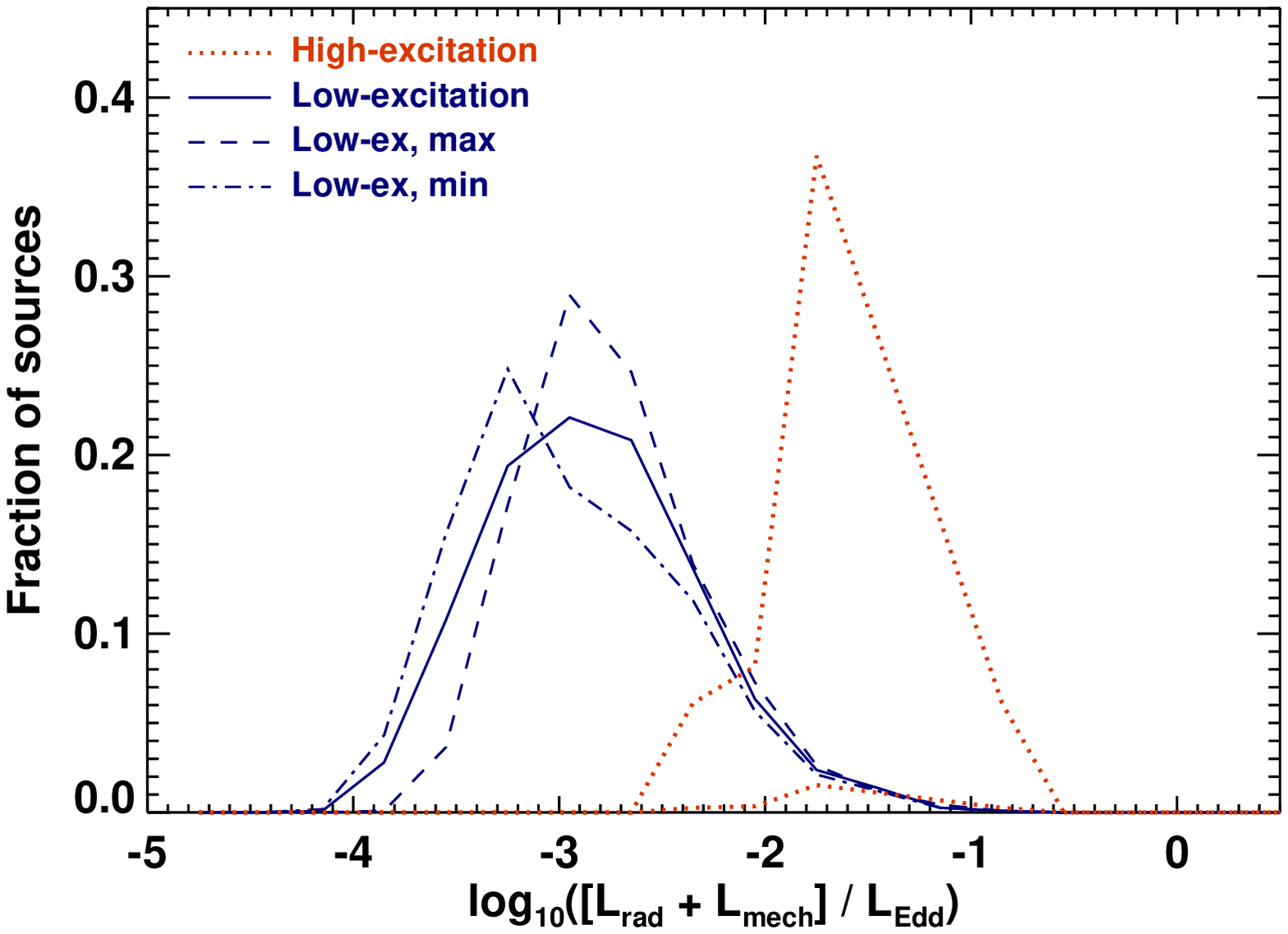,width=8.6cm,clip=}
\vspace*{-0.1cm}
}
\centerline{
  \psfig{file=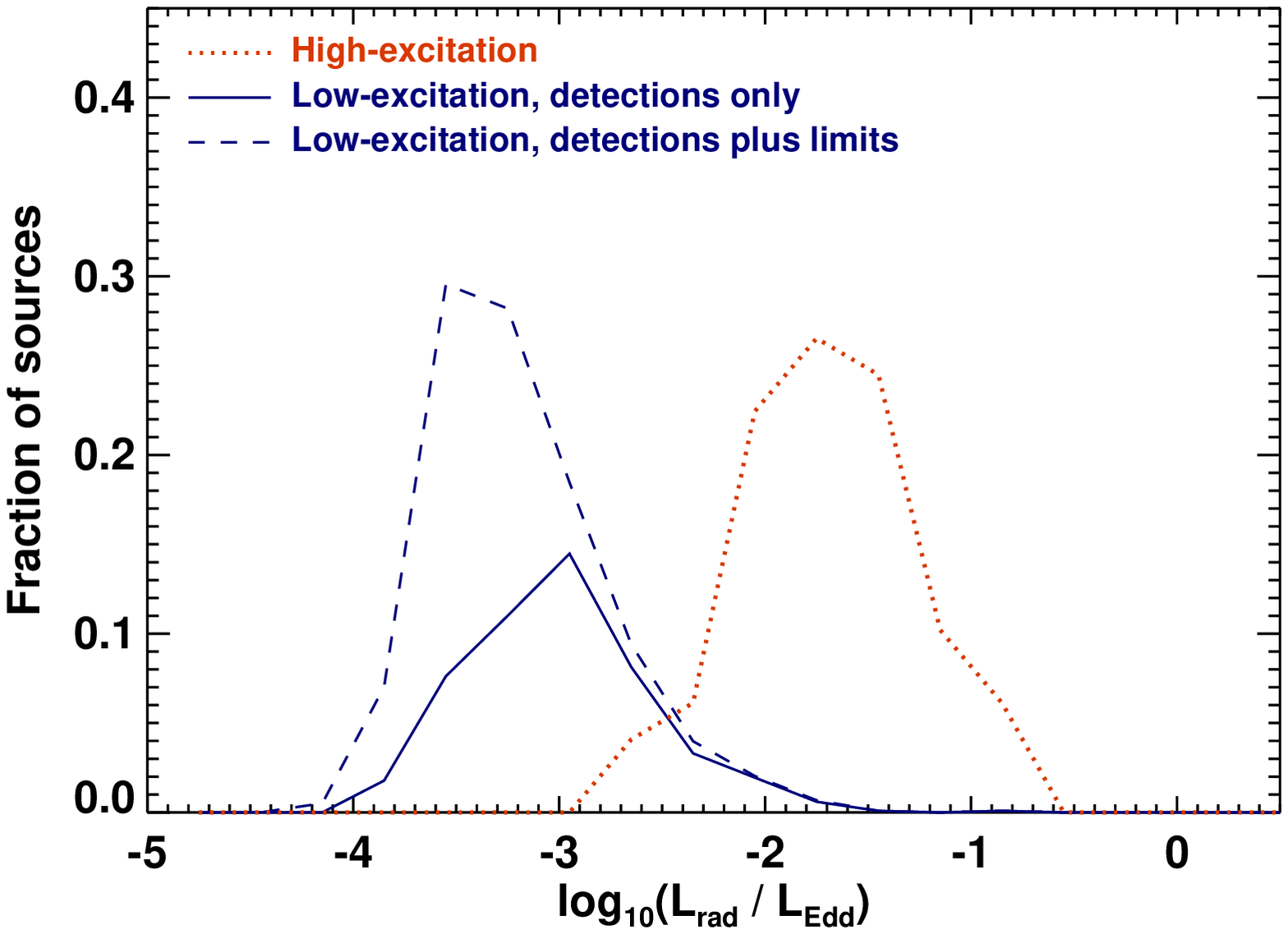,width=8.6cm,clip=}
}
\caption{\label{fig_ledd} {\it Top:} The distribution of Eddington-scaled
  accretion rates for the LERG and HERG populations separately. Analysis
  is limited to $z < 0.1$, in which redshift range essentially all radio
  sources could be robustly classified.  For the LERGs, the solid line
  shows the best-estimate distribution, using the calculated values of the
  radiative luminosity. The dashed line shows the distribution if, for all
  sources which are undetected in [OIII], the 3$\sigma$ upper limit to the
  [OIII] luminosity is used to calculate the radiative luminosity. The
  dot-dash line shows the distribution if the emission lines of all
  galaxies with H$\alpha$ equivalent width below 3\AA\ are assumed not to
  have an AGN origin.  These distributions therefore represent the allowed
  extremes of the LERG distribution, and show that the result is broadly
  similar in all three cases. For the HERGs, the dotted line is plotted
  with two different normalisations: the upper line shows the fraction of
  sources relative to the total number of HERGs, while the lower line
  shows the fraction relative to the total number of LERGs to allow direct
  comparison of numbers with the LERGs. {\it Bottom:} The distributions
  are shown considering only the radiative luminosity (ie. ignoring the
  mechanical energy in the radio jets). For LERGs, the dashed line
  indicates the maximal distribution calculated using either measurements
  of, or 3$\sigma$ upper limit to, the [OIII] luminosity. The solid line
  indicates the subset of these which are measurements rather than
  limits.}
\end{figure}

The jet mechanical luminosity was estimated from the 1.4\,GHz radio
luminosity, using the relation of \citet{cav10}, $L_{\rm mech} = 7.3
\times 10^{36} \left(L_{\rm 1.4GHz} / 10^{24}\rm{W
  Hz}^{-1}\right)^{0.70}$W. This relation was determined using the
energies associated with cavities evacuated by radio sources in the hot
X-ray gas haloes of giant ellipticals, groups and clusters of galaxies. It
is in broad agreement with minimum-energy synchrotron estimates of
\citet{wil99b}.  The scatter around the $L_{\rm mech}$ {\it vs} $L_{\rm
  1.4GHz}$ relation is observed to be about 0.7 dex \citep{cav10}.

The black hole mass of the radio source host galaxy was estimated from the
velocity dispersion of the galaxy ($\sigma_*$), as measured in the SDSS
spectrum, using the well-established $M_{BH}$-$\sigma_*$ relation.  The
determination of \citet{tre02} was adopted: log$(M_{\rm BH} / M_{\odot}) =
8.13 + 4.02 \rm{log}(\sigma_*/200$km\,s$^{-1})$. The black hole masses thus
derived define the Eddington limit for each radio source, $L_{\rm Edd} =
1.3 \times 10^{31} M_{\rm BH}/M_{\odot}$W. The intrinsic scatter in the
$M_{BH}$-$\sigma_*$ relation is less than 0.3 dex.

Combining these three relations, the Eddington-scaled accretion rate was
derived for each radio source, as $\lambda = \left(L_{\rm rad} + L_{\rm
  mech}\right) / L_{\rm Edd}$. In cases where [OIII] was undetected
(ie. signal-to-noise below 3) then the radiative luminosity was considered
to be negligible compared to the mechanical luminosity, and the
Eddington-scaled accretion rate was calculated from the mechanical
luminosity alone. For these sources, a maximum value for the
Eddington-scaled accretion rate was also calculated based on the upper
limit to the [OIII] line luminosity ($\lambda_{\rm max} = \left(L_{\rm
  rad,max} + L_{\rm mech}\right) / L_{\rm Edd}$). For sources where [OIII]
was detected, but the equivalent width of the H$\alpha$ emission line was
below 3\AA, the possibility was considered that the emission lines arise
from photo-ionisation from post-AGB stars, instead of from the AGN
\citep[see Section~\ref{sec_herglerg} and][]{cid11}. For these sources a
minimum value for the Eddington-scaled accretion rate was determined,
based only on the mechanical luminosity: $\lambda_{\rm min} = L_{\rm mech}
/ L_{\rm Edd}$. 

The upper panel of Figure~\ref{fig_ledd} shows the distribution of
Eddington-scaled accretion rates for the HERG and LERG populations
separately; analysis is limited to $z<0.1$, in which redshift range
essentially all radio sources are robustly classified. For the LERGs,
distributions are shown for each of $\lambda$, $\lambda_{\rm min}$ and
$\lambda_{\rm max}$: the three distributions are similar, demonstrating
that the high-$\lambda$ end of the distribution of Eddington-scaled
accretion rates for LERGs is robustly described, regardless of the
assumptions adopted (it should be noted that the low $\lambda$ end of the
LERG distributions is influenced by the radio selection criteria which
will lead to minimum detectable values of $L_{\rm mech}$: the true
distribution may extend to much lower values of $\lambda$ than indicated
by Figure~\ref{fig_ledd}).  The Eddington-scaled accretion rates of the
HERG and LERG populations are clearly fundamentally different: LERGs
typically display accretion rates below one per cent of the Eddington
rate, whereas HERGs typically accrete at higher Eddington rates \citep[as
  is also found for radio-quiet AGN in SDSS; cf.][]{kau09}. This result
very strongly implies that Eddington-scaled accretion rate on to the black
hole is a primary factor in determining the nature of the accretion flow.
The difference in the derived Eddington-scaled accretion rates between
HERGs and LERGs is largely driven by the estimate of accretion rates onto
the black holes (particularly the difference in $L_{\rm rad}$, calculated from
$L_{\rm OIII}$) but also in part by HERGs typically being hosted by
galaxies with less massive black holes (see Section~\ref{sec_optprops}).

For LERGs, the uncertainty in individual $\lambda$ measurements is
$\approx 0.7$ dex, dominated by the intrinsic scatter in the $L_{\rm
  mech}$ {\it vs} $L_{\rm 1.4GHz}$ relation. For HERGs, the radiative
luminosity is generally much larger than the mechanical luminosity, and so
the uncertainty in $\lambda$ measurements is $\approx 0.4$ dex. These
large uncertainties will make the observed distributions of accretion rate
broader than the intrinsic ones.  Therefore, although there is some
overlap in the accretion-rate distributions of the LERG and HERG
populations, with examples of both classes being found between a few
tenths and a few percent of Eddington, the results would be entirely
consistent with a complete dichotomy whereby the LERG/HERG classification
is entirely determined by the Eddington-scaled accretion rate
\citep[cf.][]{mer08}.  Alternatively, the overlap in accretion rates
between the two classes may imply that, although accretion rate is the
primary determinant, other factors may also be important in determining
the nature of the accretion flow on to the black hole (black hole spin is
an obvious candidate). The current dataset cannot definitively distinguish
between these possibilities, because the lack of knowledge of the precise
width and shape of the distribution of the uncertainties means that a
robust deconvolution of the distributions is not possible. It is worthy of
note, however, that if the observed LERG and HERG distributions are simply
deconvolved with Gaussians of width 0.7 and 0.4 dex, respectively, then
the two deconvolved distributions become entirely distinct, with a clean
separation occurring at 1 percent of Eddington.

Finally, the lower panel of Figure~\ref{fig_ledd} shows the distributions
considering only the radiative luminosity of the sources ($\lambda_{\rm
  rad} = L_{\rm rad} / L_{\rm Edd}$), ie. ignoring the mechanical
luminosity of the radio jets. This is a somewhat unphysical comparison
since it ignores the major energetic output of the LERGs, but is
nevertheless included for comparison with other works in the literature
which have defined the Eddington ratio in this way. The division between
the two populations is even more clear in this analysis.

\section{The nature of the radio source host galaxies}
\label{sec_optprops}

Differences in host galaxy properties between the HERG and the LERG
populations may provide interesting insight in to either the triggering
mechanism of the radio activity, or the effect of the radio source on its
host galaxy. Early work with relatively small samples compared radio
galaxies with strong emission lines against those with weak or absent
lines, and suggested that the strong emission line sources were less
luminous, with lower velocity dispersions, bluer colours and lower
mass-to-light ratios \citep[ie. younger stellar
  populations;][]{smi89,smi90b}. These results have been confirmed with
much larger radio source samples by \citet{kau08}, who studied the
differences between SDSS-selected radio sources with and without
detectable emission lines, and found the host galaxies of emission line
radio sources to have lower stellar masses, lower velocity dispersions,
lower 4000\AA\ break strengths, and stronger Balmer absorption features
than those without emission lines. However, it is important to note that
many LERGs do display emission lines, and so these results should be
tested with properly defined LERG and HERG samples.

\begin{figure*}
\centerline{
\begin{tabular}{cc}
\psfig{file=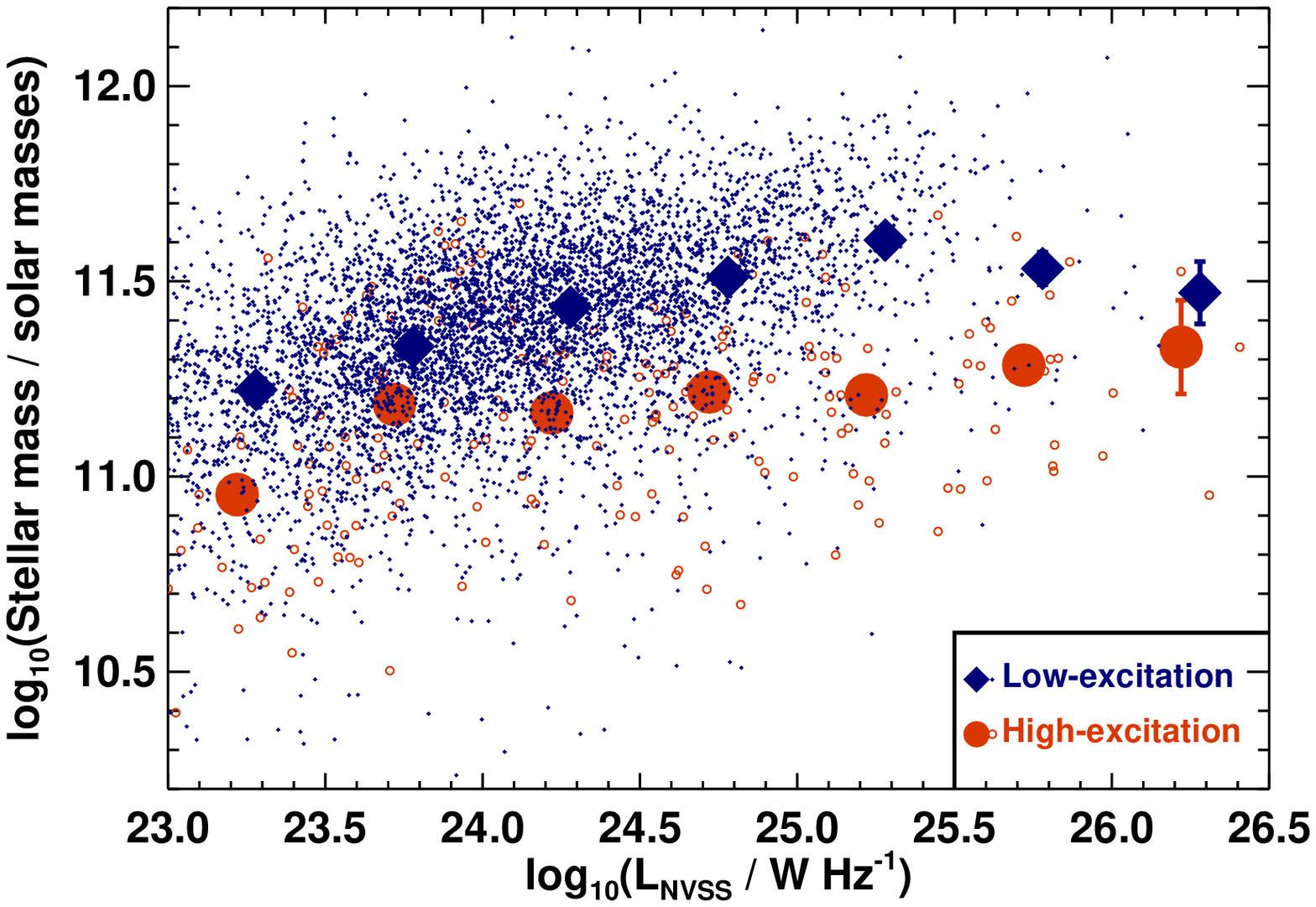,width=8.4cm,clip=} &
\psfig{file=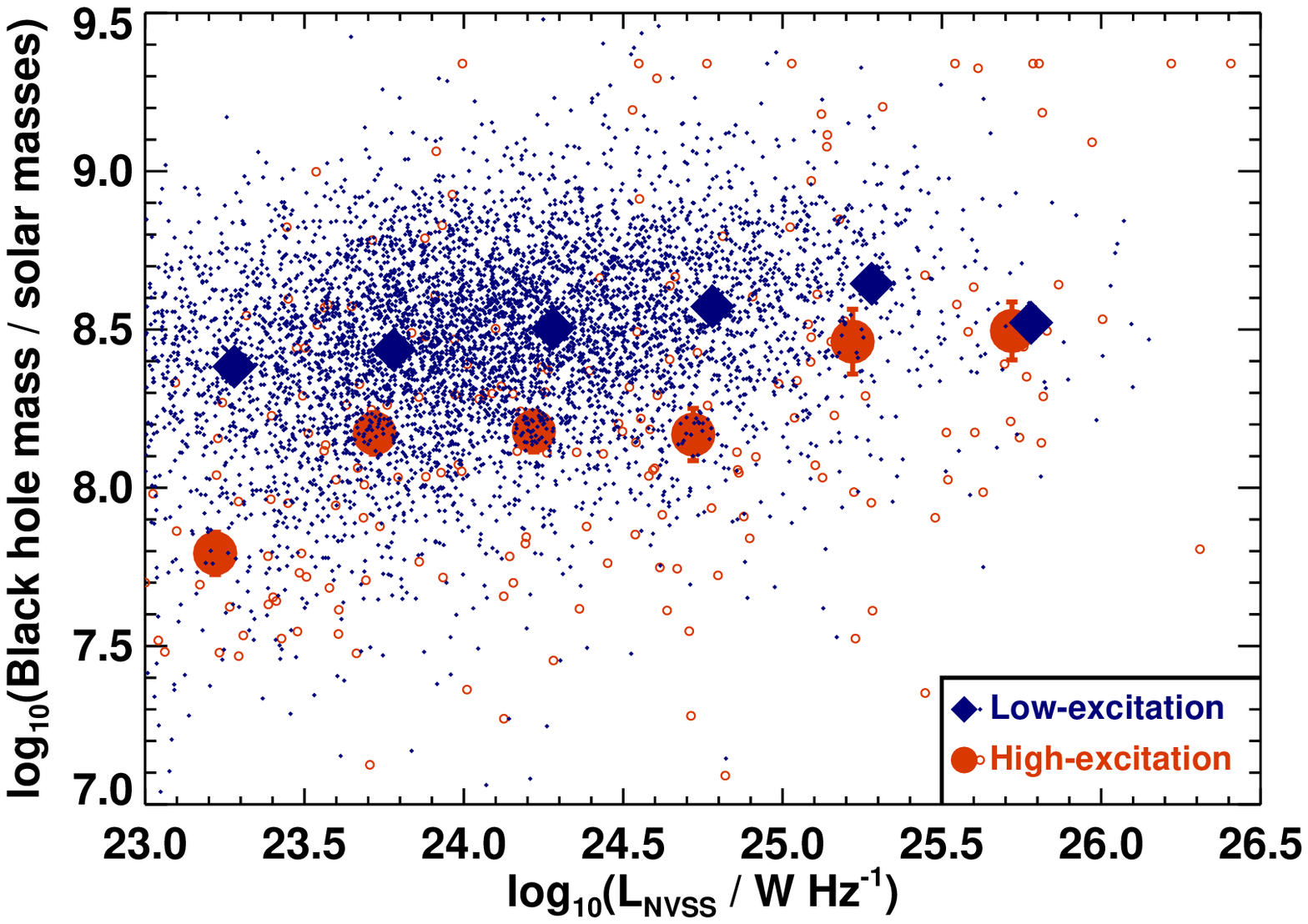,width=8.4cm,clip=} \\
\psfig{file=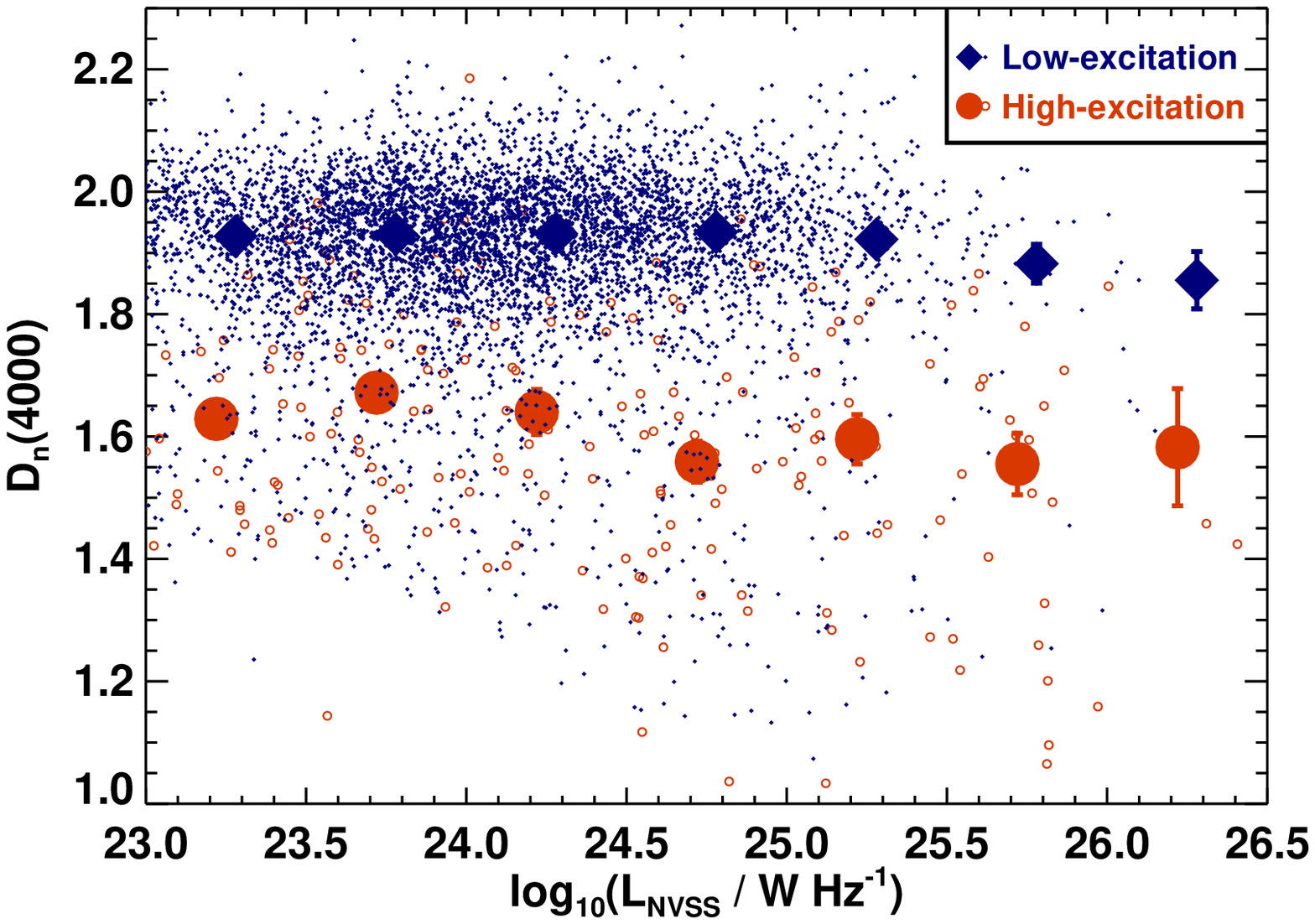,width=8.4cm,clip=} &
\psfig{file=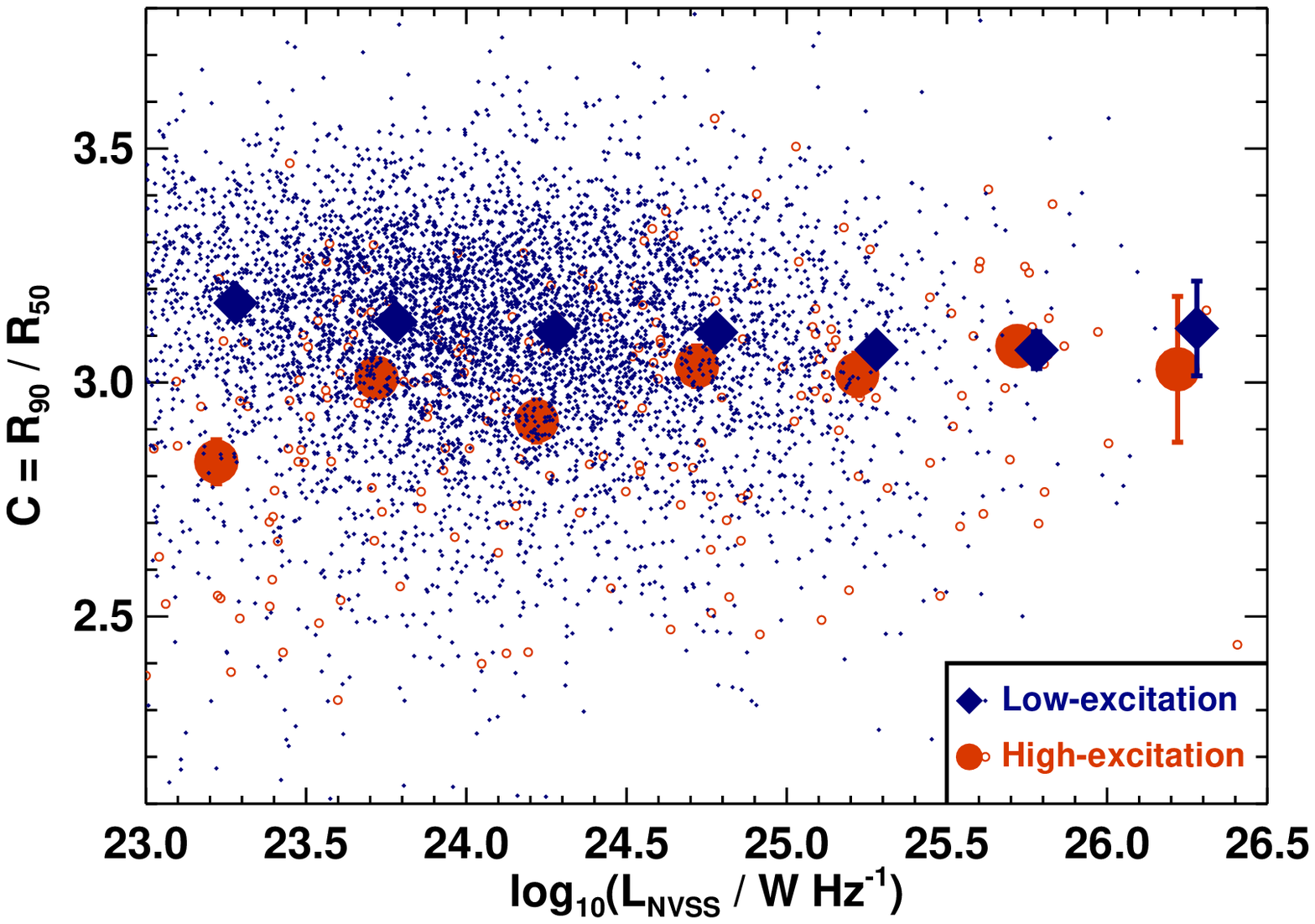,width=8.4cm,clip=}
\end{tabular}
}
\caption{\label{fig_props} The distributions of stellar mass (upper left),
  black hole mass (upper right), 4000\AA\ break strength (lower left) and
  concentration index ($C=R_{90}/R_{50}$, where $R_{90}$ and $R_{50}$ are
  the radii containing 90\% and 50\% of the light, respectively; lower
  right), as a function of radio luminosity, for the LERG and HERG
  populations separately. The larger symbols indicate the mean values for
  each population in radio luminosity bins.}
\end{figure*}

\citet{lin10} also studied large samples of radio galaxies selected from
SDSS, in their case separated into different radio morphological types
\citep[Fanaroff and Riley Class 1 and 2 sources -- FR1/2;
][]{fan74}. These two radio morphological classes show a broad overlap
with the LERG and HERG classes, respectively, and therefore many
differences between LERGs and HERGs may also be reflected by differences
between FR1s and FR2s. For this reason, many previous analyses of host
galaxy properties, accretion rates, and luminosity functions have
concentrated on differences between the FR1/2 classes \citep[e.g.][and
  references therein]{led96,ghi01,cao04,rig08,gen10}. However, there are
substantial differences between the HERG/LERG and FR1/2 segregations,
since a significant population of LERG FR2s exists \citep{lai94}. Indeed,
\citet{lin10} found the most significant differences when they compared
the host galaxy properties of the most edge-brightened FR2s with strong
emission lines against those of the other radio sources: this is closer to
a HERG/LERG split. They found this FR2 subset to be hosted by lower mass
galaxies, live in sparser environments \citep[cf.][]{pre88,smi90}, and
have higher accretion rates than the rest of the radio source
population. Again this argues for the need to investigate clean LERG and
HERG samples.

\begin{figure*}
\centerline{
\psfig{file=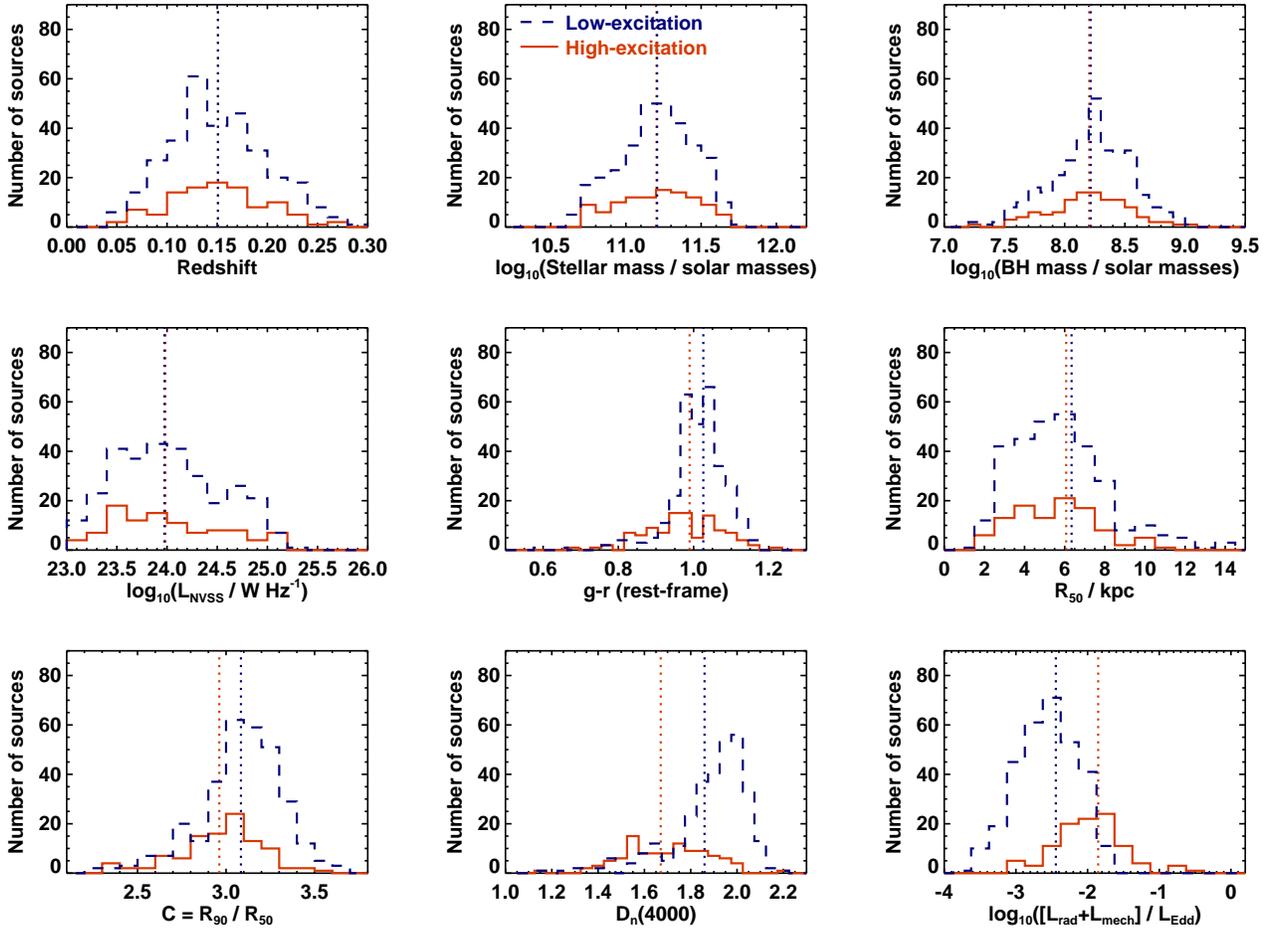,angle=90,width=17.5cm,clip=}
}
\caption{\label{fig_matched} Histograms showing the distribution of radio
  source and host galaxy properties for samples of LERGs and HERGs which
  have been matched in redshift, stellar mass, black hole mass, and radio
  luminosity (with three LERGs matched to each HERG). The first four
  panels indicate the success of this matching, with the distributions of
  LERGs and HERGs agreeing well in these four parameters. The remaining
  panels show differences between the HERG and LERG populations in galaxy
  colours, sizes, concentration indices, 4000\AA\ break strengths, and
  Eddington-scaled accretion rates. Vertical dotted lines indicate the
  mean values for each population.}
\end{figure*}

Figure~\ref{fig_props} shows the distributions of stellar mass, black hole
mass, 4000\AA\ break strength, and concentration index ($C=R_{90}/R_{50}$,
where $R_{90}$ and $R_{50}$ are the radii containing 90\% and 50\% of the
light in the r-band) for the host galaxies of the LERG and HERG
populations, as a function of radio luminosity\footnote{Note that the same
  result for concentration index is found if an upper redshift limit of
  $z=0.15$ is applied, indicating that any effects of seeing on the
  measurement of $C$ are unimportant.}. It is immediately apparent that
the HERG selection picks out host galaxies which are less massive and have
lower black hole masses than those of the LERGs, in line with the results
of \citet{kau08} and \citet{lin10}, and earlier works. The 4000\AA\ breaks
strengths of the HERGs are also lower than those of the LERGs at all radio
luminosities, indicating younger stellar populations \citep[note that
  there is no issue of AGN light contamination; see discussion
  in][]{kau08}.  At low radio luminosities the concentration indices of
the HERGs are also lower.

Many properties of galaxies in the local Universe correlate strongly with
the stellar mass. The difference in 4000\AA\ breaks strength between LERGs
and HERGs in Figure~\ref{fig_props} cannot therefore be properly
interpreted until it is known whether it is simply driven by the lower
typical stellar mass of the HERGs. To address this, matched samples of
LERGs and HERGs were created. For each HERG, a search was made for LERGs
which were matched to $\pm 0.02$ in $z$, $\pm 0.1$ in log$M$, $\pm 0.1$ in
log$M_{\rm BH}$ and $\pm 0.25$ in log$L_{\rm NVSS}$. If at least three
such LERGs were found, then three of these were randomly selected for the
matched sample. If three matches could not be found then the HERG was
excluded from the analysis.

Figure~\ref{fig_matched} shows histograms of a variety of host galaxy
properties for the matched LERG and HERG samples. The first four panels
illustrate the distributions of redshift, stellar mass, black hole mass
and radio luminosity for the matched samples, and demonstrate the success
of the matching. The remaining five panels show the distributions of
galaxy colour ($g-r$), size ($R_{50}$ in kpc), concentration index,
4000\AA\ break strength and Eddington-scaled accretion rate for the
matched LERGs and HERGs. HERGs are seen to be bluer, smaller, less
concentrated, and have lower 4000\AA\ breaks than LERGs of the same
stellar and black hole mass and radio luminosity. The offset in galaxy
sizes is of marginal significance, as calculated by KS-tests on 1000
iterations of the random matching selection, but the offsets of the other
four parameters are each significant at $> 99.9$\% significance level. The
difference in accretion rates is still extremely strong in the matched
samples (note that the median of the LERG distribution is higher than in
Figure~\ref{fig_ledd} because the matching with the HERGs pushes the LERG sample
towards higher radio luminosities and lower black hole masses). The bluer
colours and lower 4000\AA\ breaks are consistent with the HERGs being
associated with ongoing star formation activity, as is also seen in
similar radio-quiet AGN \citep[e.g.][]{kau03c}. The smaller average sizes
and lower concentration indices may be related to the triggering mechanism
of the sources. In particular, if LERGs are often fuelled by accretion
from their hot gas haloes, then they might be more preferentially located
at the centres of groups or clusters \citep[cf.][]{kau08,lin10}, where
host galaxies are typically larger and may have more extended (cD-type)
light profiles.

\section{Summary and Interpretation}
\label{sec_concs}

A large sample of radio sources drawn from the SDSS has been classified
into high- and low-excitation radio galaxies, and the nature of these two
different populations has been investigated. The main results are:

\begin{itemize}
\item Local radio luminosity functions have been derived separately for
  the LERGs and HERGs for the first time. Both populations are found
  across the full range of radio luminosities studied, although LERGs
  dominate the population at low radio luminosities and HERGs at high
  luminosities. The two populations appear to be switching in dominance at
  $L_{\rm 1.4 GHz} \sim 10^{26}$W\,Hz$^{-1}$.

\item HERGs show evidence for strong cosmic evolution at all radio
  luminosities, while LERGs are consistent with little or no evolution. 
  This difference, coupled with the changing population mix as a function 
  of radio luminosity, helps to drive the strong luminosity-dependence
  of the evolution seen in the overall radio luminosity function.

\item The accretion rates of the HERGs and LERGs are fundamentally
  different. HERGs typically have accretion rates between one and ten
  percent of Eddington. LERGs, in contrast, predominantly accrete at rates
  of below one percent of Eddington.

\item HERGs are hosted by galaxies of lower mass and lower black hole mass
  than LERGs of the same radio luminosity. The host galaxies of HERGs are
  also bluer and have lower 4000\AA\ break strengths than LERGs of the
  same mass and radio luminosity, indicating the presence of associated
  star formation. LERGs are larger and have more extended light profiles
  than HERGs, consistent with the interpretation that they are more likely
  to be hosted by central galaxies of groups and clusters.
\end{itemize}

These results are consistent with the developing picture of radio-loud
AGN, in which HERGs are fuelled at relatively high rates in
radiatively-efficient standard accretion disks by cold gas, perhaps
brought in through mergers and interactions, and with some of the cold gas
leading to associated star formation. The requirement for significant cold
gas supplies means that these sources are much more prevalent at earlier
cosmic epochs where merger rates and gas fractions were larger. In
contrast, LERGs are fuelled at relative low rates, through
radiatively-inefficient accretion flows, largely by gas associated with
the hot X-ray haloes surrounding the galaxy or its group or cluster
(although gas from any other source fuelling the black hole at low
accretion rates would also lead to a LERG). Regardless of whether the
accretion occurs directly from the hot gas through the Bondi mechanism, or
after the hot gas has cooled, this gas source allows the setting up of an
AGN feedback cycle, since this is the gas directly affected by any
radio-AGN activity \citep[cf.][]{bes06a}. In this picture, LERGs will be
associated with massive galaxies, which have old passive stellar
populations. Given that massive galaxies show little evolution out to $z
\sim 1$, the lack of cosmic evolution seen in the LERG population is as
expected.

One of the most interesting results coming out of this study is the almost
distinct nature of the accretion rate properties of the LERG and HERG
classes. This is in line with earlier indications comparing FR1 and FR2
sources \citep{ghi01}, and with more recent results for a switch in
accretion rate between BL Lacs and flat-spectrum radio quasars
\citep{wu11,ghi11}. This result fits in well with theoretical calculations
of accretion modes onto black holes \citep{nar95}, and is in line with
what is found in galactic X-ray binary systems. X-ray binaries display
three well-defined spectral states \citep[e.g.][and references
  therein]{fen04}: a `low/hard' state in which the source is characterised
by hard X-ray emission, low-power radio jets are ubiquitous, and the radio
and X-ray luminosities are correlated \citep[][possible analogue to the
  radiatively inefficient AGN state]{gal03}; a `high/soft' state dominated
by a thermal X-ray component characteristic of a standard thin accretion
disk (possible analogue to radio-quiet quasar-like AGN); a short-lived
`intermediate' or `transition' state between the two, where both the
thermal accretion disk and powerful radio jets are seen (possible analogue
to radio-loud quasar-like AGN).  The switch between these different
accretion states in X-ray binary systems has been shown to depend upon the
accretion rate onto the black hole, with the switch occurring at between 1
and a few percent of the Eddington rate \citep{mac03} - very similar to
the value derived in this paper for radio-AGN.

\citet{kor06} demonstrated that spectral states analogous to those of
X-ray binaries could be identified in local AGN, while other analogies
between the black holes in X-ray binaries and those in AGN have been
uncovered through examinations of the relationships between radio and
X-ray luminosities and black hole mass \citep[the so-called `Fundamental
  plane of black hole activity';][]{mer03,fal04}. On the basis on those
results, and the assumption that not only do AGN show these same three
spectral states but that there is also the same dependence on accretion
rate between them, synthesis models of AGN have been constructed
\citep[e.g.][]{mer08}. The results in this paper offer strong evidence in
support of these AGN synthesis models, and the broad principles of the
analogy of AGN with X-ray binary systems, by both confirming previous
indications of a lower limit to the accretion rate of radiatively
efficient AGN, and demonstrating that an upper limit of around a percent
Eddington applies to the radiatively-inefficient population.

\section*{Acknowledgements} 

PNB is grateful for financial support from the Leverhulme Trust.  The
research makes use of the SDSS Archive, funding for the creation and
distribution of which was provided by the Alfred P. Sloan Foundation, the
Participating Institutions, the National Aeronautics and Space
Administration, the National Science Foundation, the U.S. Department of
Energy, the Japanese Monbukagakusho, and the Max Planck Society.  The
research uses the NVSS and FIRST radio surveys, carried out using the
National Radio Astronomy Observatory Very Large Array: NRAO is operated by
Associated Universities Inc., under co-operative agreement with the
National Science Foundation. The authors would like to thank Guinevere
Kauffmann, Andrea Merloni and Martin Hardcastle, amongst others, for
interesting discussions on this topic, and an anonymous referee for helpful
suggestions.

\bibliography{pnb} 
\bibliographystyle{mn2e}

\appendix

\section{Separation of star-forming galaxies and radio-loud AGN}
\label{app_sfagn}

This appendix provides details of the techniques used to classify the
selected SDSS radio sources as either star-forming galaxies or radio-loud
AGN. It is important to clarify that the attempt made here is to classify
the origin of the radio emission, and hence that radio-quiet AGN are
classified together with the star-forming galaxies, rather than
contaminating the radio-loud AGN category.

\begin{figure}
\centerline{
\psfig{file=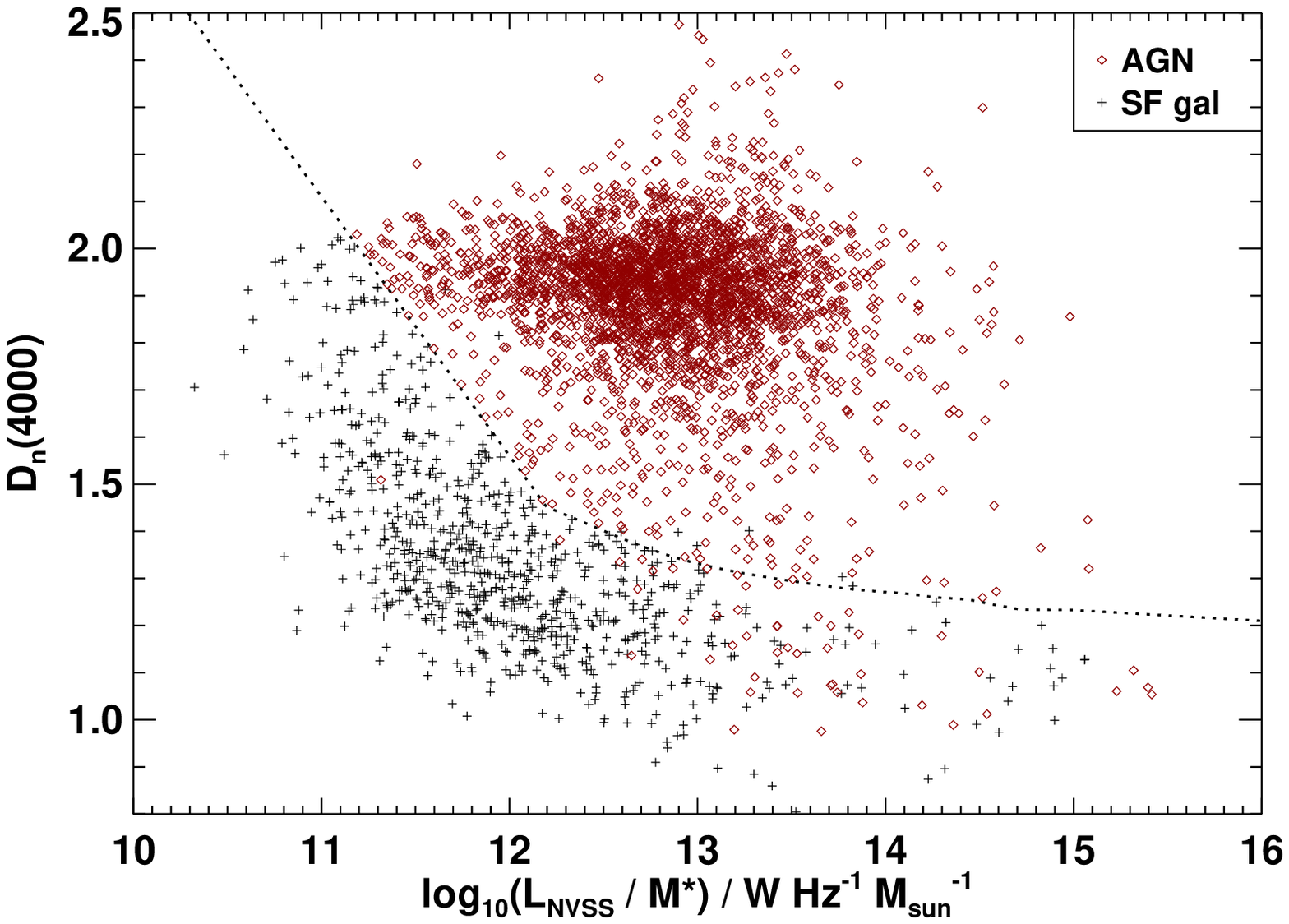,width=8.8cm,clip=} 
}
\centerline{
\psfig{file=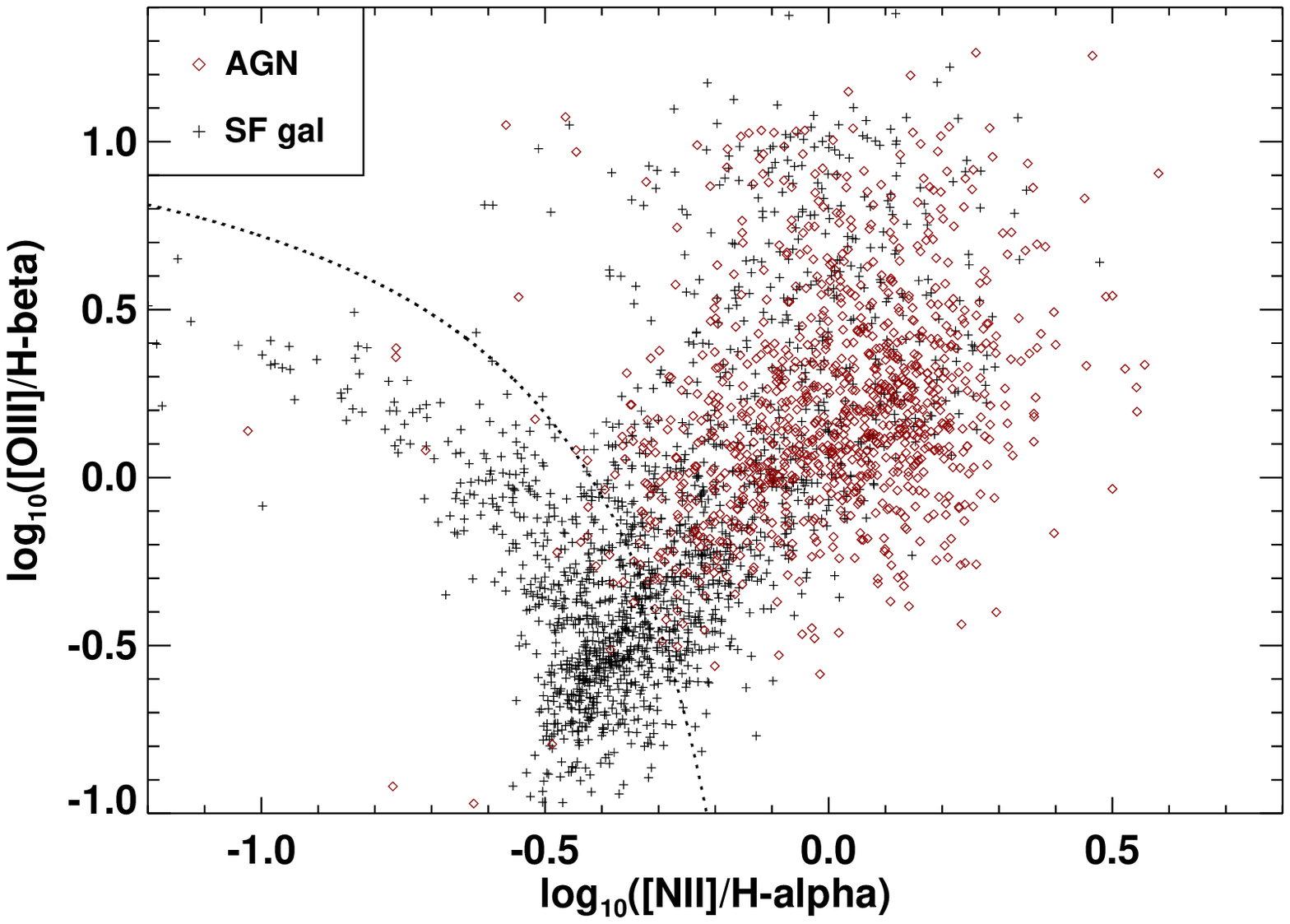,width=8.8cm,clip=} 
}
\centerline{
\psfig{file=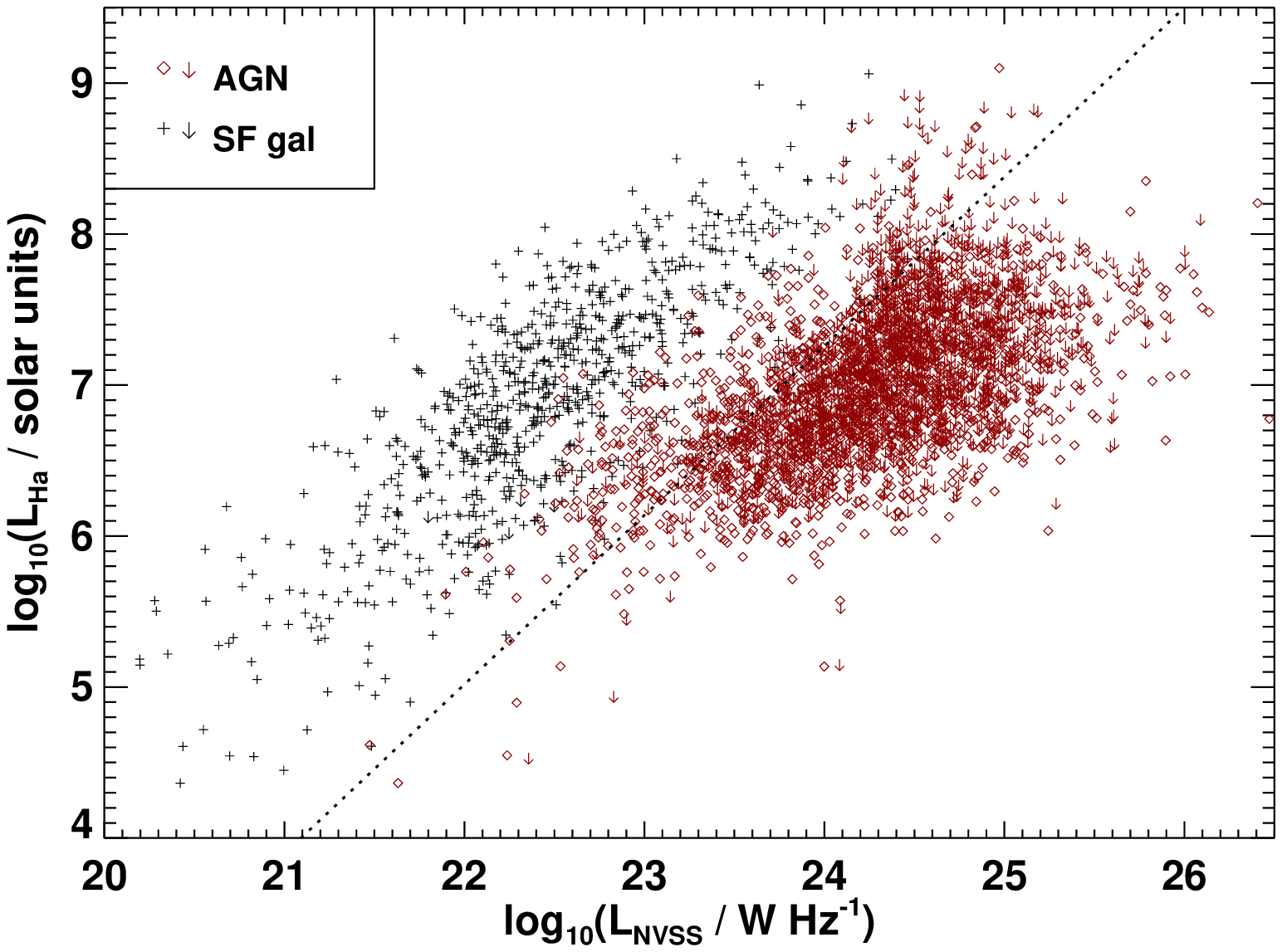,width=8.8cm,clip=} 
}
\caption{\label{fig_sfagn} The location of radio sources on the three
  classification lines used to separate radio-loud AGN from galaxies where
  the radio emission is powered by star-formation. The top plot is the
  `D$_{4000}$ {\it vs} $L_{rad}$/M' method, developed by
  \citet{bes05a}. The middle plot shows the widely-used `BPT' emission
  line ratio diagnostic. The lower plot shows the relationship between
  H$\alpha$ and radio luminosity. In all plots, the dotted lines indicate
  the division used for that classification method. Sources plotted as red
  diamonds are classified as radio-loud AGN in the overall classification,
  while star-forming galaxies appear as black crosses. In the lower plot,
  arrows indicate upper limits to the H$\alpha$ luminosity.}
\end{figure}

Three mechanisms for separating the two classes of sources are considered:
(i) using the relationship between the 4000\AA\ break strength and the
ratio of radio luminosity per stellar mass \citep[cf.][]{bes05a},
hereafter referred to as the `D$_{4000}$ {\it vs} $L_{\rm rad}$/M' method;
(ii) using emission line diagnostics, in particular the ratio of [O{\sc
    iii}]~5007 and H$\beta$ line fluxes, and that of [N{\sc ii}]~6584 and
H$\alpha$ \citep[cf.][]{bal81}, hereafter referred to as the `BPT' method;
(iii) using the relation between the H$\alpha$ emission line luminosity
and the radio luminosity -- the `$L_{H\alpha}$ {\it vs} $L_{rad}$'
method. The precise divisions adopted for each of these three methods are
described in Section~\ref{sec_sfagn1} and are illustrated in
Figure~\ref{fig_sfagn}. Following this, the manner in which the results of
these methods are combined to produce an overall classification is
explained in Section~\ref{sec_sfagn2}.

\begin{table*}
\caption{\label{tab_sfagn} The table presents the numbers of sources in
  each combination of classifications by the three different
  classification methods, together with the overall classification adopted
  and the rationale for that classification. It should be emphasised that
  `AGN' in the classification refers to sources classified as radio-loud
  AGN, whereas `SF' refers to sources where the radio emission is powered
  by star-formation (albeit that a radio-quiet AGN may also be present).}
\begin{tabular}{cccccl}
\hline
 D$_{4000}$ {\it vs} & BPT & L$_{H\alpha}$ & No of & Overall & Rationale for overall classification \\
 $L_{\rm rad}$/M & & {\it vs} L$_{\rm rad}$ & Sources & class  & \\
\hline
AGN & AGN & AGN & 1078 & AGN  & Unambiguous AGN \\
AGN & AGN & ??  & 0    & --   & No sources \\
AGN & AGN & SF  & 847  & AGN  & Mostly close to cut line in L$_{H\alpha}$ but clear AGN in other plots \\
AGN & ??  & AGN & 8053 & AGN  & All weak-lined, reliable AGN  \\
AGN & ??  & ??  & 1938 & AGN  & Weak-lined AGN. Majority high-z. L$_{H\alpha}$ limit nearly allows AGN classification \\
AGN & ??  & SF  & 222  & AGN  & Clear AGN in D$_{\rm 4000}$. Close to L$_{H\alpha}$ cut; offset from bulk of SF population \\
AGN & SF  & AGN &  25  & AGN  & Radio excess. Near BPT cut line. Probably radio-loud AGN with associated SF \\
AGN & SF  & ??  & 0    & --   & No sources \\
AGN & SF  & SF  & 40   & SF   & Close to D$_{\rm 4000}$ cut line. Most have high extinction. \\
??  & AGN & AGN & 42   & AGN  & Clear AGN; almost all have high L$_{\rm rad}$ and high $D_{4000}$ but no mass estimate. \\
??  & AGN & ??  & 0    & --   & No sources \\
??  & AGN & SF  & 55   & SF   & Most likely scenario is radio-quiet AGN with associated SF \\
??  & ??  & AGN & 1166 & AGN  & Mostly $z>0.3$. Almost all have high L$_{\rm rad}$ and high $D_{4000}$ but no mass estimate. \\
??  & ??  & ??  & 1767 & AGN  & All but six have $z>0.3$ and $L_{\rm rad} > 10^{25} W Hz^{-1}$  \\
??  & ??  & SF  & 12   &SF$^*$& Properties mostly appear like SF galaxies; see footnote \\
??  & SF  & AGN & 0    & --   & No sources \\
??  & SF  & ??  & 0    &  --  & No sources \\
??  & SF  & SF  & 29   &SF$^*$& Consistent classification, but see footnote \\
SF  & AGN & AGN & 37   & AGN  & Mostly high z, high L$_{\rm rad}$. Probably radio-loud AGN with SF activity \\
SF  & AGN & ??  & 0    & --   & No sources \\
SF  & AGN & SF  & 1582 &SF$^*$& Mostly low luminosity; almost certainly radio-quiet AGN \\
SF  & ??  & AGN & 73   & AGN  & Mostly high z, high L$_{\rm rad}$. Probably radio-loud AGN with SF activity \\
SF  & ??  & ??  & 42   &SF$^*$& Clear SF in D$_{\rm 4000}$ but too distant for emission line classifications \\
SF  & ??  & SF  & 35   &  SF  & Consistent classification \\
SF  & SF  & AGN & 19   &SF$^*$& Mostly low L$_{\rm rad}$ SF galaxies near L$_{H\alpha}$ cut line, but see footnote.\\
SF  & SF  & ??  & 0    &  --  & No sources \\
SF  & SF  & SF  & 1224 &  SF  & Unambiguous SF \\
\hline
\end{tabular}
\parbox{\textwidth}{$^*$ In these classes, there exist small populations of objects with
$L_{\rm rad} > 10^{24.5}$W\,Hz$^{-1}$ and $z>0.3$. The properties of these
subsets of high $L_{\rm rad}$, high-$z$ galaxies appear much more like
radio-loud AGN. Therefore, although the bulk of these objects are
classified overall as SF galaxies, the subsets of $L_{\rm rad} >
10^{24.5}$W\,Hz$^{-1}$ and $z>0.3$ sources are classified as radio-loud
AGN.}
\end{table*}

\subsection{The three classification methods}
\label{sec_sfagn1}

The `D$_{4000}$ {\it vs} $L_{rad}$/M' method was developed by
\citet{bes05a}, on the basis that star-forming galaxies with a wide range
of star formation histories occupy a similar locus in this plane. This is
because, for star-forming galaxies, both $L_{rad}$/M and D$_{4000}$ depend
broadly on the specific star formation rate of the galaxy. Radio-loud AGN
will have enhanced values of $L_{rad}$ and are thus separable on this
plane. \citet{bes05a} adopted a division line of 0.225 higher in
D$_{4000}$ than the track produced by a galaxy with an exponentially
declining star formation rate of 3\,Gyr e-folding time. They demonstrated
that this method was generally very successful by comparison with other
methods used in the literature. Further work by \citet{kau08} indicated
that this selection line may be somewhat too shallow at low values of
$L_{\rm rad}$/M, misclassifying some star-forming galaxies as AGN. Taking
account of this, the division line of \citet{bes05a} was modified such
that for $L_{\rm rad}$/M $< 12.2$, a straight-line cut is adopted with
equation $D_{4000} = 1.45 - 0.55 (L_{\rm rad}/M - 12.2)$.  Most radio
sources are classifiable by this method, but $\approx 17$\% of sources
are not, due to the lack of a mass estimate in the MPA-JHU SDSS
value-added catalogues. These unclassifiable sources are almost always
the highest redshift sources, $z>0.3$, selected in samples other than the
`main galaxy sample' (ie. with $r>17.77$), although small numbers of lower
redshift sources where the automated algorithm had failed are also present.

The `BPT' method has been widely used by a number of authors to divide
star-forming galaxies from AGN, since the different spectrum of the
ionising radiation leads to different emission line ratios.  As outlined
by \citet{bes05a}, however, there are problems relating to its direct use
for identification of a clean radio-loud AGN sample. In particular, it is
well-established that AGN and star-formation activity are closely related
\citep[e.g.][]{kau03c}.  Radio-quiet AGN may therefore be detected in the
radio on the basis of their on-going star-formation activity and
identified as AGN on the basis of their emission line ratios, thus
contaminating the radio-loud sample. Nevertheless, the emission line
ratios can offer useful information in many cases. Here, the division
proposed by \citet{kau03c} is adopted, namely that galaxies with
log([O{\sc iii}]/H$\beta$) $>$ 1.3 + 0.61 / (log([N{\sc ii}]/H$\alpha$) -
0.05) are classified as AGN. Just under 30\% of the radio source sample
are classifiable by this method, but the rest lack detections (or
definitive limits) for at least one line.

The `$L_{H\alpha}$ {\it vs} $L_{rad}$' method is based on the premise that,
for star-forming galaxies, both the H$\alpha$ luminosity and the radio
luminosity provide a direct measure of the star formation rate and these
properties are therefore correlated. For radio-loud AGN, a far higher
radio luminosity to H$\alpha$ luminosity ratio is observed, and so the
location of galaxies in the L$_{H\alpha}$ {\it vs} L$_{\rm rad}$ plane can
be used to identify radio-loud AGN \citep[cf.][]{kau08}. Ideally the
extinction-corrected H$\alpha$ luminosity would be used, since dust
attenuation can decrease the H$\alpha$--to--radio luminosity ratio of
star-forming galaxies. However, this is only possible for the subset of
galaxies for which H$\beta$ is also detected, to allow a dust attenuation
estimate from the Balmer decrement, and even then the attenuation
correction that should be used is different for star-forming galaxies and
AGN. Here the observed H$\alpha$ luminosity is therefore used instead,
with an adopted division line of log$(L_{H\alpha} / L_{\odot}) = 1.12
\times ({\rm log} (L_{rad} / {\rm W Hz}^{-1}) - 17.5)$. This cut is designed to
be conservative, in the sense that galaxies classified as radio-loud AGN
by this cut should be secure AGN, whereas the SF class may be contaminated
by some AGN close to the cut line. Almost 80\% of the radio sources are
classifiable by this method, either using the observed H$\alpha$
luminosity, or because the upper limit on that luminosity allowed clear
classification.

\subsection{Combination into a final classification}
\label{sec_sfagn2}

For each of the three classification methods, every radio source is
classified as a radio-loud AGN, classified as having its radio emission
associated with star formation, or is unclassified. There are therefore 27
possible combinations of classification. Table~\ref{tab_sfagn} indicates
the number of galaxies in each of these 27 different classes and the
resulting classification adopted. These overall classifications are
arrived at by examining in detail the properties of the galaxies in each
class, especially where different classification methods disagree. For
example, in most cases where objects miss the D$_{4000}$ {\it vs} $L_{\rm
  rad}$/M classification due to the lack of a stellar mass estimate, the
values of D$_{4000}$ and $L_{\rm rad}$ are both so high that the galaxy
will be classified as an AGN for any plausable value of the stellar mass
and so that classification can be securely adopted. Similarly, in some
classes with disputed classifications, if the sources lie very clearly
within the AGN regime in two of the plots but just on the SF side of the
cut line in the third, then that adds weight to an overall AGN
classification. The rationale for the final classification is
included in Table~\ref{tab_sfagn}.

Figure~\ref{fig_rlfsfagn} shows the radio luminosity functions derived for
star-forming galaxies and radio-loud AGN. The smooth nature of the derived
luminosity functions, especially at low and high luminosities where small
numbers of misclassifications would have pronounced effects, offers broad
support to the success of the classification scheme, as does the
comparison with previous determinations from \citet{mac00}, \citet{sad02}
and \citet{mau07}. The slight offset of the AGN line from previous
determinations below a radio luminosity of $L \approx 10^{23}$W\,Hz$^{-1}$
may be caused by the different classification of radio-quiet AGN, which
here are classified along with the star-forming galaxies. In any case,
objects of these low luminosities are excluded from most of the analyses
in this paper.

\label{lastpage}
\end{document}